\documentclass[%
 reprint,
 superscriptaddress,
 amsmath,amssymb,
 aps,
]{revtex4-1}

\usepackage{graphicx}
\usepackage{dcolumn}
\usepackage{bm}
\usepackage{xcolor}
\usepackage{url}

\usepackage{subfigure}
\usepackage{multirow}

\begin{document}


\title{An analysis of polarized parton distribution functions with nonlinear QCD evolution equations}

\author{Chengdong Han}
\affiliation{Institute of Modern Physics, Chinese Academy of Sciences, Lanzhou 730000, China}
\affiliation{School of Nuclear Science and Technology, University of Chinese Academy of Sciences, Beijing 100049, China}

\author{Gang Xie}
\affiliation{Institute of Modern Physics, Chinese Academy of Sciences, Lanzhou 730000, China}
\affiliation{Guangdong Provincial Key Laboratory of Nuclear Science, Institute of Quantum Matter, South China Normal University, Guangzhou 510006, China}

\author{Rong Wang}
\email{rwang@impcas.ac.cn}
\affiliation{Institute of Modern Physics, Chinese Academy of Sciences, Lanzhou 730000, China}
\affiliation{School of Nuclear Science and Technology, University of Chinese Academy of Sciences, Beijing 100049, China}

\author{Xurong Chen}
\email{xchen@impcas.ac.cn}
\affiliation{Institute of Modern Physics, Chinese Academy of Sciences, Lanzhou 730000, China}
\affiliation{School of Nuclear Science and Technology, University of Chinese Academy of Sciences, Beijing 100049, China}
\affiliation{Guangdong Provincial Key Laboratory of Nuclear Science, Institute of Quantum Matter, South China Normal University, Guangzhou 510006, China}


\date{\today}

\begin{abstract}
We present the polarized parton distribution functions from a QCD analysis
of the worldwide polarized deep inelastic scattering data,
based on the dynamical parton distribution model.
All the sea quarks and gluons are dynamically generated from QCD radiations,
with the nonperturbative input contains only the polarized valence quark distributions.
This approach leads to a simple parametrization,
which has only four free parameters.
In the analysis, we apply the DGLAP equations with parton-parton recombination corrections.
The parameterized nonperturbative input at an extremely low $Q_0^2$
reproduces well the spin-dependent structure functions measured at high $Q^{2}$.
Comparisons with experimental observations and some other polarized parton distribution functions are also shown.
Our results are in good agreement with the experimental data and consistent with some other parameterized models.
Furthermore, our analysis gives the positive polarized gluon distribution
and it suggests that the gluon distribution plays an important role to the proton spin content.
The polarized antiquark distributions are non-zero at high $Q^2$
but quite small compared to polarized valence quark distribution,
based on this dynamical parton model analysis.
This analysis shows smaller statistical uncertainties for the polarized sea quark and gluon distributions,
thanks to the fewer free parameters used for the parametrization of the initial polarized PDFs.
\end{abstract}

\maketitle


\section{Introduction}
\label{sec:intro}

In the quark model, the nucleon spin comes from the sum of the spins
of the three constituent quarks.
However the famous European-Muon-Collaboration (EMC) experiment \cite{EuropeanMuon:1987isl,EuropeanMuon:1989yki}
showed that the quarks carry only a small fraction of the nucleon spin,
which brought us a question on where the whole nucleon spin originates from.
This long-standing problem is called the ``proton spin puzzle''
and it indicates that the nucleon structure is more complicated
than the quark model description.
At a more fundamental level, the missing spin content can be attributed
to the orbital angular momenta of quarks and gluons,
and the helicity of the gluons, of which the mechanism should be
explained with the modern quantum chromodynamic (QCD) theory.

In experiments, the nucleon spin structure is mainly probed
by the polarized lepton-nucleon deep inelastic scattering (DIS) process.
After decades of efforts in theoretical calculations and experimental measurements,
our understanding on the nucleon spin structure has been significantly improved,
especially in the valence quark region.
At the same time, measurements of the orbital angular momentum,
the transversity distributions, and the helicity distributions
of sea quarks and gluons are still quite limited.
The current polarized DIS experiments on proton, deuteron and $^3$He targets 
mainly provide the spin asymmetry $A_{1}$ or the spin structure function $g_{1}$
\cite{EuropeanMuon:1987isl,EuropeanMuon:1989yki,HERMES:1998cbu,HERMES:2006jyl,COMPASS:2010wkz,
COMPASS:2015mhb,E143:1998hbs,SpinMuon:1998eqa,CLAS:2017qga,COMPASS:2005xxc,COMPASS:2016jwv,
E142:1996thl,HERMES:1997hjr,E154:1997xfa,JeffersonLabHallA:2004tea,Kramer:2005qe}.
Thanks to the factorization theory, the polarized parton distribution functions (PDFs)
are determined in the global analysis of these high-energy scattering data.
Many groups have studied this kind of QCD analysis \cite{Gehrmann:1995ag,deFlorian:1994qn,Gluck:1995yr,
Gluck:2000dy,Bartelski:1995hw,Bourrely:2001du,Leader:1999rh,Leader:2001kh,
Blumlein:2002qeu,Hirai:2003pm,Leader:2005ci,deFlorian:2005mw,Bourrely:2005kw,
Hirai:2006sr,deFlorian:2009vb,Khorramian:2010qa,Nocera:2014gqa,Nematollahi:2021ynm,Adamiak:2021ppq},
with the polarized PDFs obtained, such as GRSV \cite{Gluck:2000dy}, BB \cite{Blumlein:2002qeu}, AAC \cite{Hirai:2003pm},
LSS \cite{Leader:2005ci}, DSSV \cite{deFlorian:2009vb}, KATAO \cite{Khorramian:2010qa}, NNPDF \cite{Nocera:2014gqa} and JAM \cite{Adamiak:2021ppq}.
It is amazing to see that the polarized valence quark distributions
$\Delta u_{v}(x, Q^{2})$ and $\Delta d_{v}(x, Q^{2})$
are well determined from these analyses of the polarized DIS data.

However the polarized antiquark distributions $\Delta \bar{q}(x, Q^{2})$ and the polarized gluon distribution
have not been well constrained by the analyses with the current experimental data.
The polarized gluon distribution has great uncertainty
and we still do not understand the gluon helicity contributes how much the nucleon spin content.
The flavor-dependence of polarized sea quark distributions is also not that clear.
To quantify the spin contents of various partons inside the proton,
the first moments of polarized parton distribution functions are usually calculated.

In this work, we perform a QCD analysis under the theoretical framework
of the dynamical parton distribution assumption \cite{Gluck:1977ah,Wang:2016sfq}.
The polarized sea quark and gluon distributions are
purely generated from the QCD evolution, which are called the dynamical parton distributions.
In this model, there are completely no intrinsic sea quarks
and gluons at the very low initial scale $Q^{2}_{0}$.
These dynamical parton distributions will provide the important knowledge
on polarized sea quark and gluon distributions.
For parton distribution evolution starting from extremely low initial scale $Q_0^2\sim 0.1$ GeV$^2$ in this analysis,
we adopt the state-of-the-art strong coupling $\alpha_{\rm s}$,
which is saturated instead of going up to infinity in the infrared region \cite{Binosi:2016nme,Rodriguez-Quintero:2018wma,Cui:2019dwv}.
Since the partons overlap greatly at low resolution scale,
we adopt the Dokshitzer-Gribov-Lipatov-Altarelli-Parisi (DGLAP) equations with parton recombination corrections
\cite{Wang:2016sfq,Gribov:1984tu,Mueller:1985wy,Chen:2013nga}.
At low $Q^2$, the correlations between the partons slow down the parton splitting processes.
With these theoretical tools the polarized PDFs are obtained.

The organization of the paper is as follows.
Section \ref{sec:exp-data} lists the current available polarized DIS data we used for the analysis.
In Sec. \ref{sec:nonperturbative-input}, we show the polarized nonperturbative input
for the dynamical parton distribution model.
In Sec. \ref{sec:evolution-eqs}, we discuss the polarized QCD evolution equations
with parton-parton recombination corrections
and the saturated strong running coupling constant $\alpha_{\rm s}$.
In Sec. \ref{sec:analysis-method}, we discuss the least-square analysis
and the Hessian matrix method for the PDF uncertainties.
Section \ref{sec:results} presents our analysis results,
the comparisons with the experimental measurements
and some other polarized PDF analyses.
Finally, a brief summary is given in Sec. \ref{sec:summary}.

\section{Experimental data}
\label{sec:exp-data}

The proton, neutron and deuteron spin structure functions ($g_{1}^{p}(x,Q^2)$, $g_{1}^{n}(x,Q^2)$ and $g_{1}^{d}(x,Q^2)$)
measured in DIS process provide an excellent set of data to study the polarized PDFs.
This type of experiment requires both the polarized lepton beam and the polarized target.
In this global QCD analysis, the experimental data of $g_{1}^{p}$ and $g_{1}^{d}$ we used are the followings.
The EMC collaboration obtained the spin-dependent structure function $g_{1}^{p}$ over a wide $x$ range ($0.01 < x < 0.7$)
with the longitudinally polarized muon beams \cite{EuropeanMuon:1987isl,EuropeanMuon:1989yki};
The SMC collaboration at CERN also obtained the spin structure functions of the proton and the deuteron
with the scattering of high-energy polarized muon \cite{SpinMuon:1998eqa};
The E143 \cite{E143:1998hbs} experiment at SLAC gave the proton and deuteron spin structure functions
$g_{1}^{p}$, $g_{1}^{d}$, with the polarized electron beams at the energies of 29.1 GeV, 16.2 GeV, and 9.7 GeV;
The HERMES experiment at DESY exploits the DIS process of the 27.6 GeV longitudinally polarized positrons
bombarding on the longitudinally polarized hydrogen and deuterium gas targets \cite{HERMES:1998cbu, HERMES:2006jyl};
The COMPASS05 and COMPASS17 experiments at CERN \cite{COMPASS:2005xxc, COMPASS:2016jwv} presented
the spin-dependent structure function $g_{1}^{d}$ of the deuteron in the kinematical range
of $1<Q^{2}<100$ GeV$^{2}$ and $0.004<x<0.7$ with a 160 GeV polarised muon beam
and a polarized $^{6}$LiD target.
The COMPASS10 and COMPASS16 experiments are with the polarized muons of 200 GeV which are scattered
by the longitudinally polarized solid-state NH$_{3}$ target \cite{COMPASS:2010wkz, COMPASS:2015mhb};
The CLAS collaboration at Jefferson Laboratory collected the data with a longitudinally polarized electron beam
at energies from 1.6 GeV to 5.7 GeV on the longitudinally polarized proton
(NH$_{3}$) and deuteron (ND$_{3}$) targets \cite{CLAS:2017qga}
over a range of $Q^{2}$ from 0.05 to 5.0 GeV$^{2}$.

In addition, the extraction of neutron spin structure function $g^{n}_{1}(x,Q^2)$ is usually
measured through DIS of polarized electrons by a polarized $^{3}$He target as a quasi polarized neutron target.
For the neutron structure function, we used the following experimental data.
The E142 \cite{E142:1996thl} experiment at SLAC studied the neutron spin structure functions $g^{n}_{1}$
from DIS of polarized electrons by a polarized $^{3}$He target at incident energies of 19.42, 22.66 and 25.51 GeV.
The HERMES experiment (HERMES97) \cite{HERMES:1997hjr} at HERA reported the measurement of the neutron
spin structure functions $g^{n}_{1}$ in DIS using 27.5 GeV longitudinally polarized positrons
incident on a polarized $^{3}$He internal gas target.
The E154 collaboration \cite{E154:1997xfa} reported on a precision measurement of the neutron spin structure function $g^{n}_{1}$ using
DIS of polarized electron on polarized $^{3}$He with the kinematical range of $0.014<x<0.7$ and $1<Q^{2}<17$ GeV$^{2}$.
The Jefferson Lab Hall A Collaboration (Jlab04) \cite{JeffersonLabHallA:2004tea} reported on measurements of the neutron polarized
structure functions $g^{n}_{1,2}$ in the valence quark region, with $Q^{2}$ = 2.7, 3.5 and 4.8 GeV$^{2}$.
The Jlab05 presented the first measurement of the $Q^{2}$-dependence of the neutron spin structure function $g^{n}_{2}$
at low $Q^{2}$ with longitudinally polarized electrons that were scattered off a
polarized $^{3}$He target in Hall A at Jefferson Lab \cite{Kramer:2005qe}.

\begin{figure}[htbp]
\centering
\includegraphics[scale=0.4]{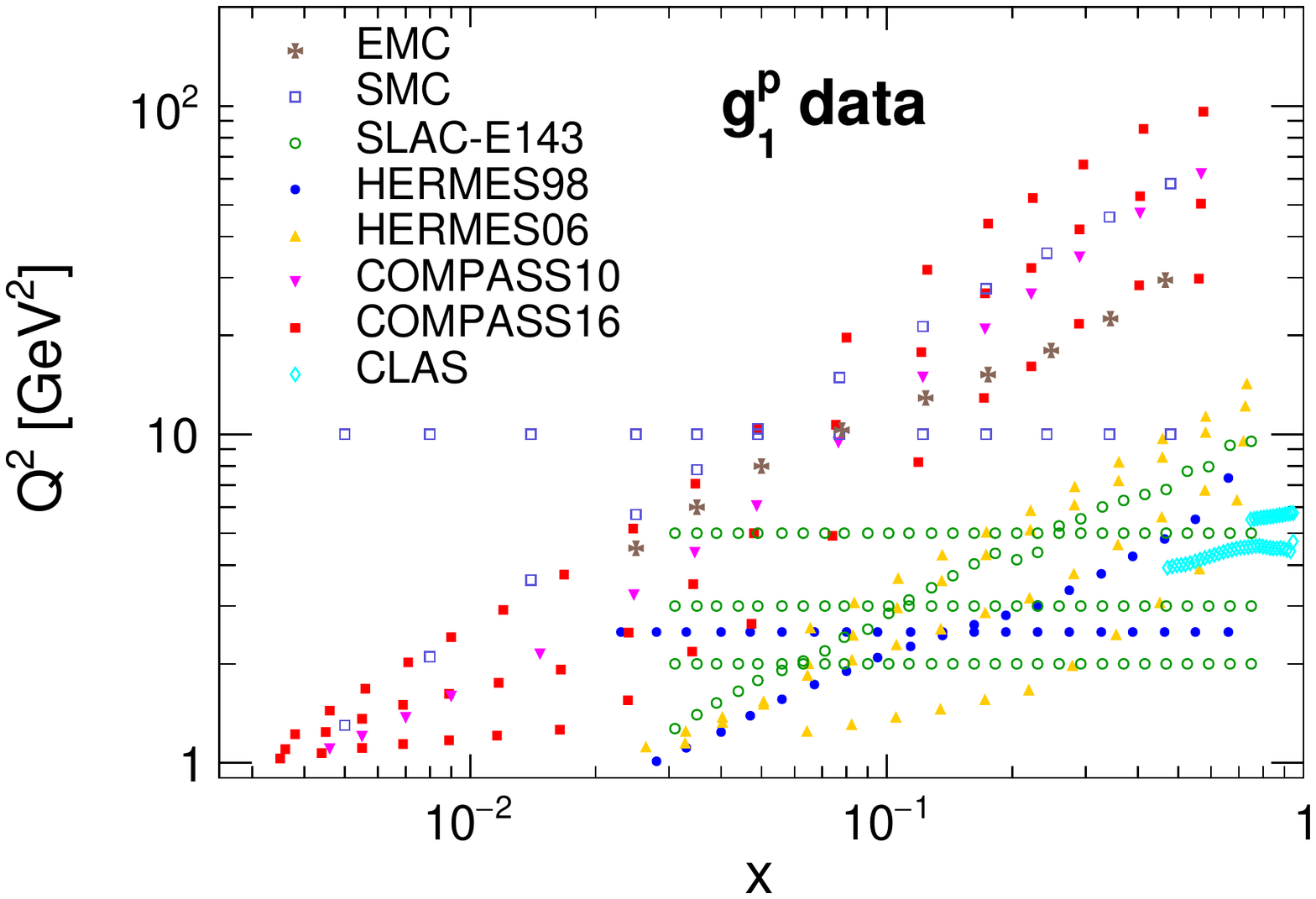}
\caption{(color online) The kinematical coverage in the ($x$, $Q^{2}$) plane of the worldwide polarized DIS $g^{p}_{1}$ data used
for our global QCD analysis.}
\label{fig:g1p-kinematic-coverage}
\end{figure}
\begin{figure}[htbp]
\centering
\includegraphics[scale=0.4]{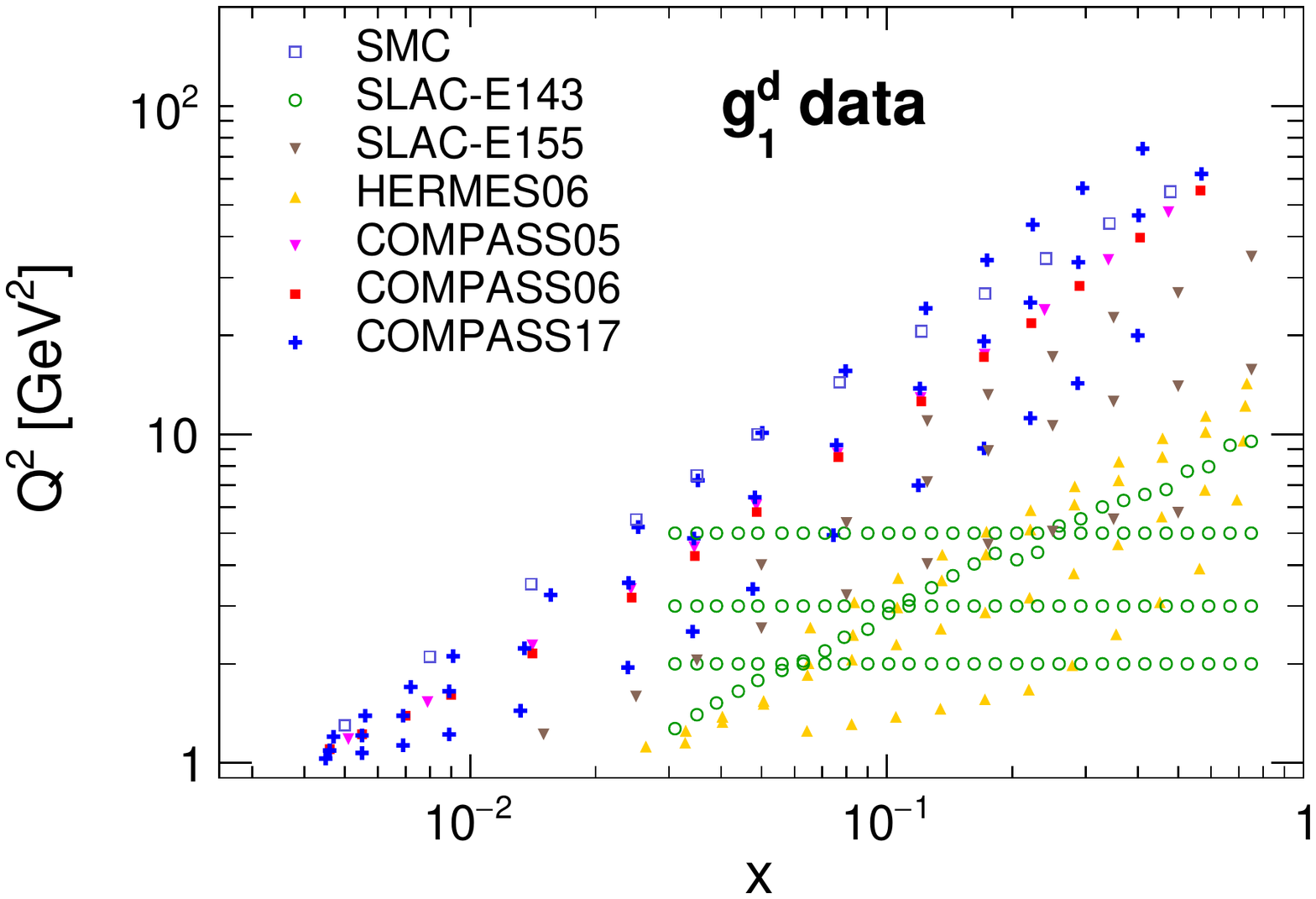}
\caption{(color online) The kinematical coverage in the ($x$, $Q^{2}$) plane of the worldwide polarized DIS $g^{d}_{1}$ data used
for our global QCD analysis.}
\label{fig:g1d-kinematic-coverage}
\end{figure}
\begin{figure}[htbp]
\centering
\includegraphics[scale=0.4]{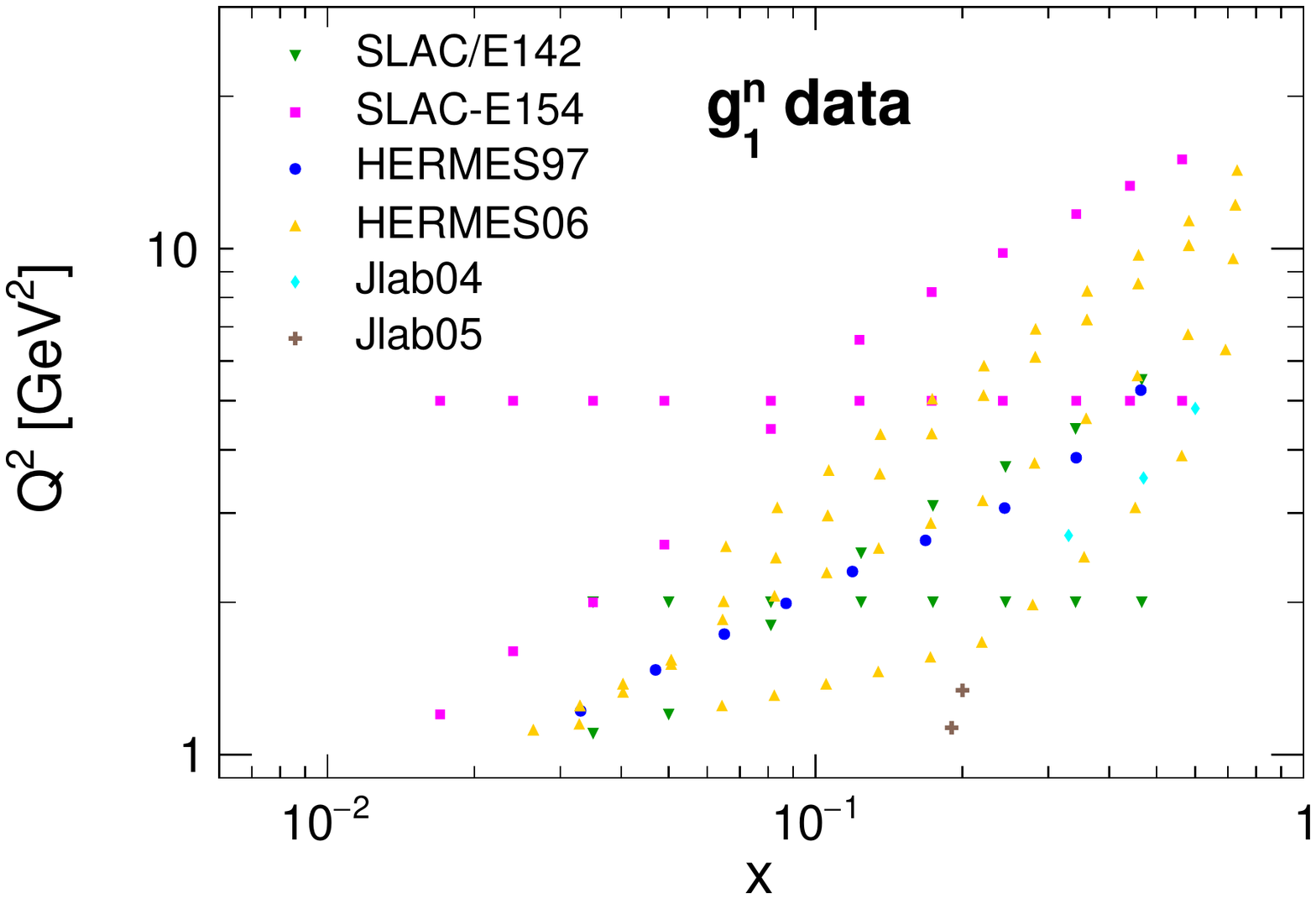}
\caption{(color online) The kinematical coverage in the ($x$, $Q^{2}$) plane of the worldwide polarized DIS $g^{n}_{1}$ data used
for our global QCD analysis.}
\label{fig:g1n-kinematic-coverage}
\end{figure}

Fig.~\ref{fig:g1p-kinematic-coverage} shows the kinematical coverage
of the available inclusive DIS g$^{p}_{1}$ data used
for our analysis of polarized PDFs.
In the figure, we see that the used experimental data points are mainly distributed in the kinematical region
of 1 GeV$^{2}$  $< Q^{2} < 100$ GeV$^{2}$ and $10^{-3} < x < 0.8$.
Fig.~\ref{fig:g1d-kinematic-coverage} shows the kinematical coverage
of the available inclusive DIS g$^{d}_{1}$ data used
for our analysis of polarized PDFs.
In the figure, we see that the used experimental data points are mainly distributed in the kinematical region
of 1 GeV$^{2}$  $< Q^{2} < 100$ GeV$^{2}$ and $10^{-3} < x < 0.8$.
Fig.~\ref{fig:g1n-kinematic-coverage} shows the kinematical coverage
of the available inclusive DIS g$^{n}_{1}$ data used
for our analysis of polarized PDFs.
In the figure, we see that the used experimental data points are mainly distributed in the kinematical region
of 1 GeV$^{2}$  $< Q^{2} < 20$ GeV$^{2}$ and $10^{-2} < x < 0.8$.
In order to better constrain the polarized antiquark distributions,
and the polarized gluon distribution from scaling violation,
more experimental data in a wider kinematical region are looking forward.
To this aim, the future facilities are proposed and under the conceptual designing,
such as the electron-ion colliders in US and China 
\cite{Accardi:2012qut,AbdulKhalek:2021gbh,Chen:2020ijn,Chen:2018wyz,Anderle:2021wcy}.

\section{Parameterized nonperturbative input and dynamical parton distribution model}
\label{sec:nonperturbative-input}

Inspired by the quark model, we know that the proton is composed of three quarks
in the first approximation.
Hence, a natural and simple assumption is that the proton consists of
only three valence quarks at the low scale $Q^{2}_{0}$.
The sea quark and gluon density distributions are all zero at the initial scale
and the sea quarks and gluons at high $Q^2$ are totally dynamically generated in QCD evolution.
Under the dynamical parton model assumption \cite{Gluck:1977ah, Vogt:1995xs, Gluck:1998xa, Jimenez-Delgado:2008orh, Jimenez-Delgado:2009may},
the gluon and sea quark distributions are well constrained by the experimental data.
Due to the large size of the partons at low resolution scale,
partons overlap more strongly at low $Q^{2}$.
Therefore in this analysis, both the spin-dependent and the spin-averaged QCD evolution equations
are implemented with parton-parton recombination corrections.

We employ $f_{+}(x, Q^{2})$ and $f_{-}(x, Q^{2})$ to refer to the parton ($f = q, \bar{q}, g$) densities
with positive and negative spin polarization respectively, which describes the partons carry a fraction $x$ of
the proton momentum at a certain resolution scale $Q^{2}$.
The difference $\Delta f(x, Q^{2}) = f_{+}(x, Q^{2}) - f_{-}(x, Q^{2})$ denotes the spin-dependent parton density of flavor $f$.
And the spin-averaged parton density is denoted as $f(x, Q^{2}) = f_{+}(x, Q^{2}) + f_{-}(x, Q^{2})$.
In this analysis of the polarized PDFs, we need two parameterized inputs for spin-dependent and spin-averaged parton densities.
The parametrization of the naive spin-averaged input as follows:
\begin{equation}
\begin{split}
xu^{V}(x, Q^{2}) = 20.2x^{1.89}(1-x)^{1.91},\\
xd^{V}(x, Q^{2}) = 7.85x^{1.25}(1-x)^{3.65},\\
x q_{i}(x, Q^{2}) = 0,\\
x g(x, Q^{2})     = 0,\\
\end{split}
\label{eq:unpolarized-input-constraints}
\end{equation}
which is taken as the result (Fit-2) of the previous work \cite{Wang:2016sfq}.
The parametrization of the spin-dependent input is written as:
\begin{equation}
\begin{split}
x\Delta u^{V}(x, Q^{2}) = Ax^{B}(1-x)^{C}, \\
x\Delta d^{V}(x, Q^{2}) = Dx^{E}(1-x)^{F}, \\
x\Delta q_{i}(x, Q^{2}) = 0, \\
x\Delta g(x, Q^{2})     = 0, \\
\end{split}
\label{eq:polarized-input-constraints}
\end{equation}
where the parameters $C$ and $F$ are fixed at 1.91 and 3.65, respectively.
The reason is that both the up polarized quark and the down polarized quark of the same flavor
have same large-$x$ behavior.

According to the quark-parton model,
the spin structure function $g_{1}^{p}$ at leading order (LO)
in the asymptotic region of high $Q^2$ is written as:
\begin{equation}
\begin{split}
  g_{1}^{p}(x, Q^{2}) = \frac{1}{2}\sum_{\rm i}e_{i}^{2}[\Delta q_{i}(x, Q^{2}) + \Delta \bar{q}_{i}(x, Q^{2})],
\end{split}
\label{eq:g1p_definition}
\end{equation}
where $i$ denotes the quark flavor, $e_{i}$ is the electrical charge of
the quark of flavor $i$ in unit of the electron charge, with $i = u, d, s$ in this work.
Considering the isospin symmetry between neutron and proton, the expression of neutron structure function $g_{1}^{n}$ at LO
can be obtained accordingly by interchanging the up and down quark PDFs.

The deuteron polarized structure function $g_{1}^{d}$ is given in terms 
of the proton and neutron structure functions as follows:
\begin{equation}
\begin{split}
  g_{1}^{d}(x, Q^{2}) = \frac{1}{2}(g_{1}^{p} + g_{1}^{n})(1 - 1.5\omega_{D}),
\end{split}
\label{eq:g1p_definition}
\end{equation}
where $\omega_{D}$ = 0.05 $\pm$ 0.01 \cite{Desplanques:1988mp} is the probability 
that the deuteron is in the D-state, given by N-N potential calculations.

\section{Polarized QCD evolution equations with parton-overlapping corrections}
\label{sec:evolution-eqs}

The DGLAP equation \cite{Gribov:1972ri,Dokshitzer:1977sg,Altarelli:1977zs}
based on the parton model and perturbative QCD theory
is an important and widely used tool to describe the $Q^{2}$-dependence of parton density.
The DGLAP equations provide a heuristic explanation for the scaling violation
with a picture of parton density variation along with the $Q^{2}$ going up.
The commonly used method to improve the accuracy of the determined PDFs is
to apply the higher-order calculations of the DGLAP equations.
Therefore many QCD evolution equations and corrections to the DGLAP equations
\cite{Gribov:1983ivg, Mueller:1985wy, Zhu:1998hg, Zhu:1999ht, Zhu:2004xj}
are under the developments.
The parton-parton recombination is among these corrections.

In this analysis, we use the parton-parton recombination effect corrected
DGLAP equations for the evolution of the unpolarized parton distributions,
for which the equations can be found in the previous paper \cite{Wang:2016sfq,Chen:2013nga}.
For the evolution of the polarized parton distributions,
we also use the DGLAP equations with parton-parton recombination corrections,
which has been proposed in the reference \cite{Zhu:2004xi,Zhu:2015uwa}.
The modified DGLAP equations for the spin-dependent parton distribution evolution
applied in this work are given by,
\begin{equation}
\begin{aligned}
Q^2\frac{dx \Delta q_v(x,Q^2)}{dQ^2}
=\frac{\alpha_s(Q^2)}{2\pi} \Delta P_{qq}\otimes \Delta q_v,
\end{aligned}
\label{eq:pDGLAP-NS}
\end{equation}
for polarized valence quark distributions,
\begin{equation}
\begin{aligned}
Q^2\frac{dx \Delta q_i(x,Q^2)}{dQ^2}
=\frac{\alpha_s(Q^2)}{2\pi}[\Delta P_{qq}\otimes \Delta q_i+\Delta P_{qg}\otimes \Delta g]\\
-\frac{\alpha_s^2(Q^2)}{4\pi R^2Q^2}\int_x^{1/2} \frac{dy}{y}x\Delta P_{gg\to q}(x,y)[yg(y,Q^2)][y\Delta g(y,Q^2)]\\
+\frac{\alpha_s^2(Q^2)}{4\pi R^2Q^2}\int_{x/2}^{x}\frac{dy}{y}x\Delta P_{gg\to q}(x,y)[yg(y,Q^2)][y\Delta g(y,Q^2)],\\
{\rm if} \quad  x\leq 1 / 2, \\
Q^2\frac{dx \Delta q_i(x,Q^2)}{dQ^2}
=\frac{\alpha_s(Q^2)}{2\pi}[\Delta P_{qq}\otimes \Delta q_i+\Delta P_{qg}\otimes \Delta g]\\
+\frac{\alpha_s^2(Q^2)}{4\pi R^2Q^2}\int_{x/2}^{1/2}\frac{dy}{y}x\Delta P_{gg\to q}(x,y)[yg(y,Q^2)][y\Delta g(y,Q^2)],\\
{\rm if} \quad  1 / 2 \leq x \leq 1,
\end{aligned}
\label{eq:DGLAP-S}
\end{equation}
for polarized sea quark distributions, and
\begin{equation}
\begin{aligned}
Q^2\frac{dx \Delta g(x,Q^2)}{dQ^2}
=\frac{\alpha_s(Q^2)}{2\pi}[\Delta P_{gq}\otimes \Delta q_i+\Delta P_{gg}\otimes \Delta g]\\
-\frac{\alpha_s^2(Q^2)}{4\pi R^2Q^2}\int_x^{1/2} \frac{dy}{y}x \Delta P_{gg\to g}(x,y)[yg(y,Q^2)][y \Delta g(y,Q^2)]\\
+\frac{\alpha_s^2(Q^2)}{4\pi R^2Q^2}\int_{x/2}^{x}\frac{dy}{y}x \Delta P_{gg\to g}(x,y)[yg(y,Q^2)][y \Delta g(y,Q^2)],\\
{\rm if} \quad  x\leq 1 / 2, \\
Q^2\frac{dx \Delta g(x,Q^2)}{dQ^2}
=\frac{\alpha_s(Q^2)}{2\pi}[\Delta P_{gq}\otimes \Delta q_i+\Delta P_{gg}\otimes \Delta g]\\
+\frac{\alpha_s^2(Q^2)}{4\pi R^2Q^2}\int_{x/2}^{1/2}\frac{dy}{y}x \Delta P_{gg\to g}(x,y)[yg(y,Q^2)][y \Delta g(y,Q^2)],\\
{\rm if} \quad  1 / 2 \leq x \leq 1,
\end{aligned}
\label{eq:DGLAP-G}
\end{equation}
for polarized gluon distribution,
where $\Delta P_{qq}$, $\Delta P_{qg}$, $\Delta P_{gq}$, $\Delta P_{gg}$ are
the standard parton splitting kernels,
and $\Delta P_{gg\to \bar{q}}$, $\Delta P_{gg\to g}$ are the gluon-gluon
recombination kernels describing the gluon overlapping effect \cite{Zhu:2015uwa}.
The factor $1/(4\pi R^2)$ in the spin-dependent evolution equations is
for the two-parton distribution normalization,
and $R$ is the correlation length of two interacting partons.
In this work, the parameter $R=3.98$ GeV$^{-1}$ is fixed as
the result of the previous global QCD analysis of the proton PDFs
with worldwide DIS data \cite{Wang:2016sfq}.

\begin{figure}[htbp]
\centering
\includegraphics[scale=0.4]{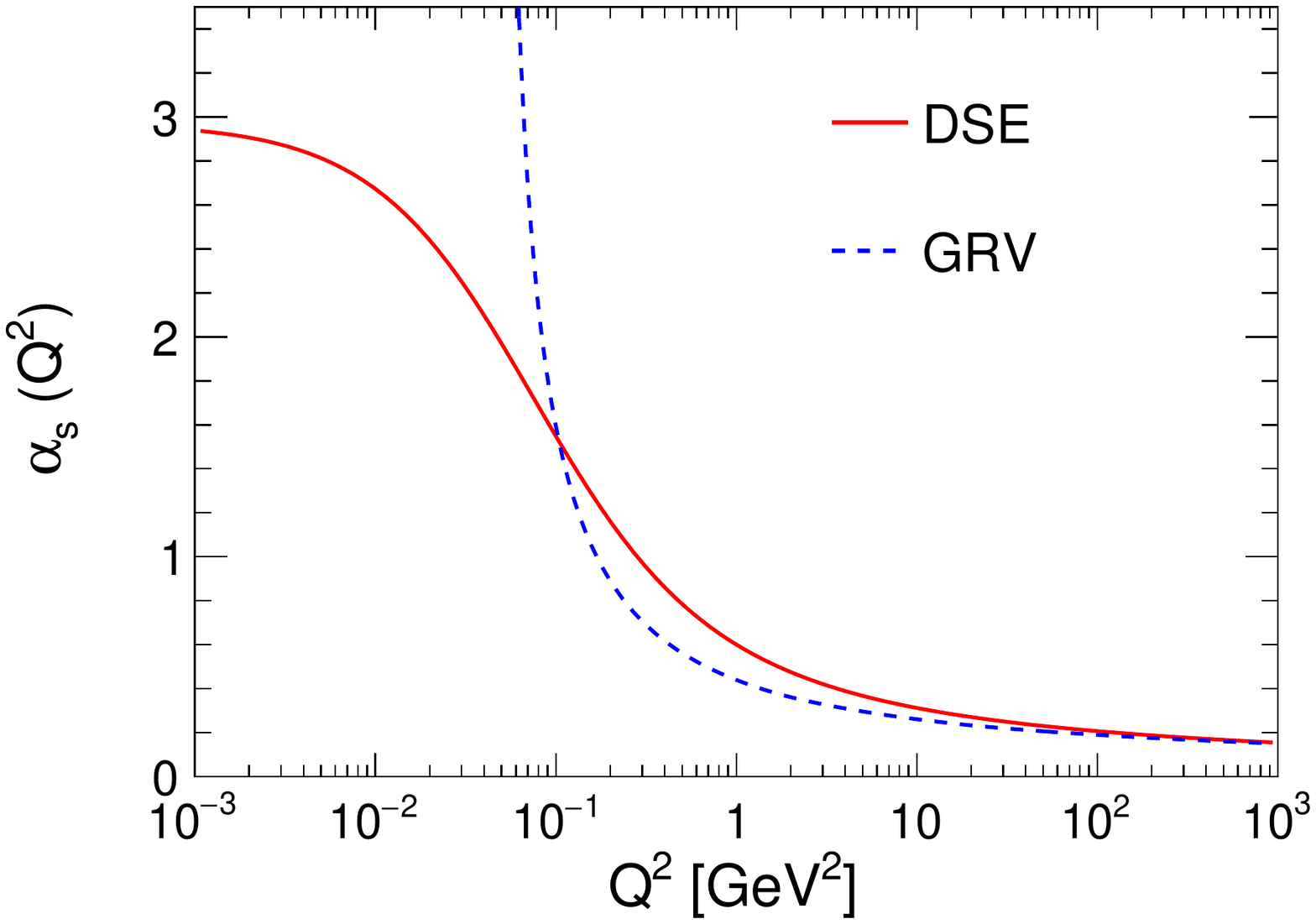}
\caption{(color online) The saturated strong running coupling from DSE 
based on the effective mass of gluon \cite{Binosi:2016nme,Rodriguez-Quintero:2018wma,Cui:2019dwv}
and the typical non-saturated strong running coupling used in GRV's QCD analysis. }
\label{fig:storng-coupling}
\end{figure}

The strong running coupling constant $\alpha_{s}$ is a fundamental parameter
for the parton evolution equations of QCD dynamics.
In this work, we employ a renormalization-group-invariant
process-independent effective strong running coupling $\alpha_{\rm s}$
fixed from lattice QCD calculation \cite{Binosi:2016nme,Rodriguez-Quintero:2018wma,Cui:2019dwv}.
It is interesting to note that at low $Q^2$ ($< 1$ GeV$^2$),
the effective strong coupling shows a saturated plateau approaching the infrared region,
and it is consistent with the experimental measurements
of Bjorken sum rule at low $Q^2$ ($<1$ GeV$^2$).

The saturated effective strong coupling \cite{Cui:2019dwv} is expressed as follows,
\begin{equation}
\begin{split}
\alpha_{\rm s}(Q^{2}) = \frac{4\pi}{\beta_{0}{\rm Ln}[(m_{\alpha}^{2} + Q^{2}) / \Lambda^{2}_{\rm QCD}]},
\end{split}
\label{eq:strong-coupling}
\end{equation}
where $\beta_{0}=(33-2n_{\rm f})/3$ refers to the one-loop $\beta$ function coefficient,
$n_{\rm f}$ is the number of flavors of the active quarks.
The $m_{\alpha}=0.43$ GeV is the effective gluon mass due to
the dynamical breaking of scale invariance \cite{Cui:2019dwv},
which is the reason why the strong interaction is saturated at the low scale.
The $\Lambda_{\rm QCD}=0.34$ GeV is the QCD cutoff for the running of strong interaction.
Fig. \ref{fig:storng-coupling} shows the strong running coupling
used in this QCD analysis, which saturates around 3 when $Q^2$ approaches zero.
Moreover, in Fig. \ref{fig:storng-coupling} we clearly see that
the saturated strong coupling is consistent with the strong coupling used in GRV analysis
in the high $Q^2$ region.

\section{Analysis method}
\label{sec:analysis-method}

To determine the polarized parton distribution functions,
only the free parameters of the parametrization of initial polarized valence quark distributions
($A, B, D, E$ in Eq. (\ref{eq:polarized-input-constraints}))
and the hadronic scale $Q_0^2$ are left to be determined,
since the nonperturbative input of spin-average parton densities are already
fixed by the previous work \cite{Wang:2016sfq}.

To get the optimal parameters for the input polarized valence distributions,
we apply the least-square regression method.
The definition of the reduced $\chi^2$ for the analysis is as follows,
\begin{equation}
\begin{split}
\chi^2 / {\rm ndf} = \frac{1}{N_{\rm tot.}-n_{\rm par.}}\sum^{N_{\rm Exp.}}_{i}
\sum_{j}^{N_{i}}\frac{(D_{j}-T_{j})^{2}}{\sigma_{j}^{2}},
\end{split}
\label{eq:chi2-def}
\end{equation}
in which $N_{\rm tot.}$ is the total number of experimental data points taken,
$n_{\rm par} = 5$ is the number of free parameters.
$N_{\rm Exp.}$ is the number of various experimental measurements used in this analysis;
$N_{i}$ is the number of data points from the $i$th experiment;
$D_{j}$ is one measured value of the experimental observables $g^{p}_{1}$, $g^{d}_{1}$, $g^{n}_{1}$;
$\sigma_{j}$ is the error for the corresponding data point $D_{j}$;
And $T_{j}$ is the corresponding theoretical prediction from QCD evolution equations.
In this work, the experimental observable $D_{j}$ we focus on is
the spin structure functions $g^{p}_{1}$, $g^{d}_{1}$, $g^{n}_{1}$ measured in DIS process.

The table \ref{tab:data_chi2value_summary} lists the obtained values of the reduced $\chi^{2}$ per dataset
and the number of polarized data points we used for this global QCD analysis.
With the least-square fitting method, the hadronic scales $Q^{2}_{0}$ and the free parameters
of the nonperturbative input are obtained,
which are listed in table \ref{tab:fit-results}.
The quality of the fit is fine with $\chi^2/ndf = 1.054$.
The experimental data points in the small-$x$ region are just a few
and the errors are quite large at present.
Objectively, the small-$x$ data for the spin structure function are highly needed,
which can be solved with the future electron-ion colliders in US and China \cite{Chen:2020ijn,Chen:2018wyz,Accardi:2012qut}.
In the analysis, the input scale is $Q^{2}_{0} = 0.068$ GeV$^{2}$ for the parameterized
nonperturbative input from the analysis of the experimental data of $g^{p}_{1}$, $g^{d}_{1}$, $g^{n}_{1}$.
This obtained input scale is consistent with the input scale
obtained from the global analysis of the experimental data of unpolarized structure function $F_2$
in the previous study \cite{Wang:2016sfq}.

\begin{figure}[htbp]
\centering
\includegraphics[scale=0.41]{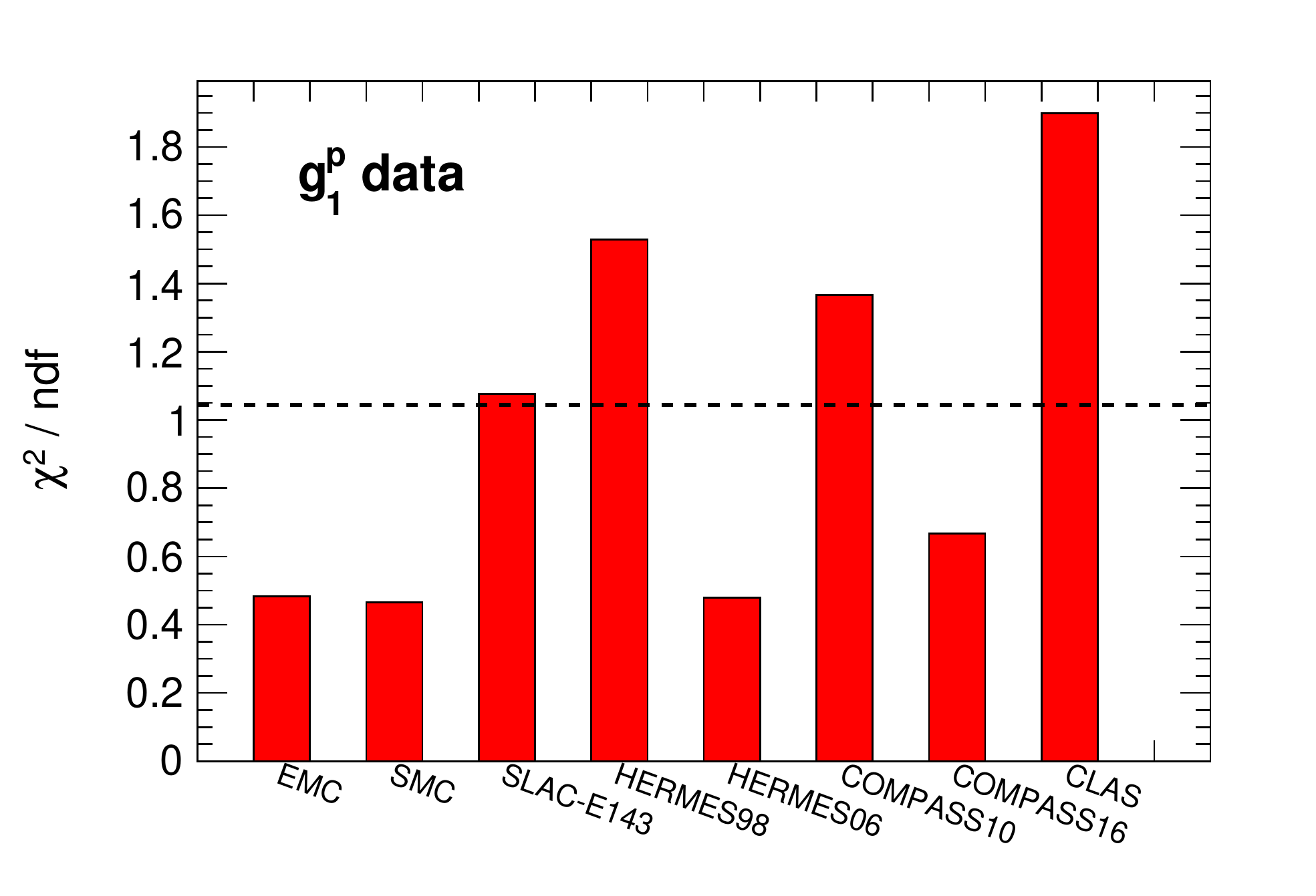}
\caption{
  (color online) Values of the reduced $\chi^{2}$ for the $g_{1}^{p}$ data of each
  experimental group (shown in Fig.\ref{fig:g1p-kinematic-coverage}), included in this global QCD analysis.
  The horizontal black dash line is the weighted average of these reduced $\chi^{2}$ over all the $g_{1}^{p}$ dataset.
}
\label{fig:g1p-chi2}
\end{figure}
\begin{figure}[htbp]
\centering
\includegraphics[scale=0.41]{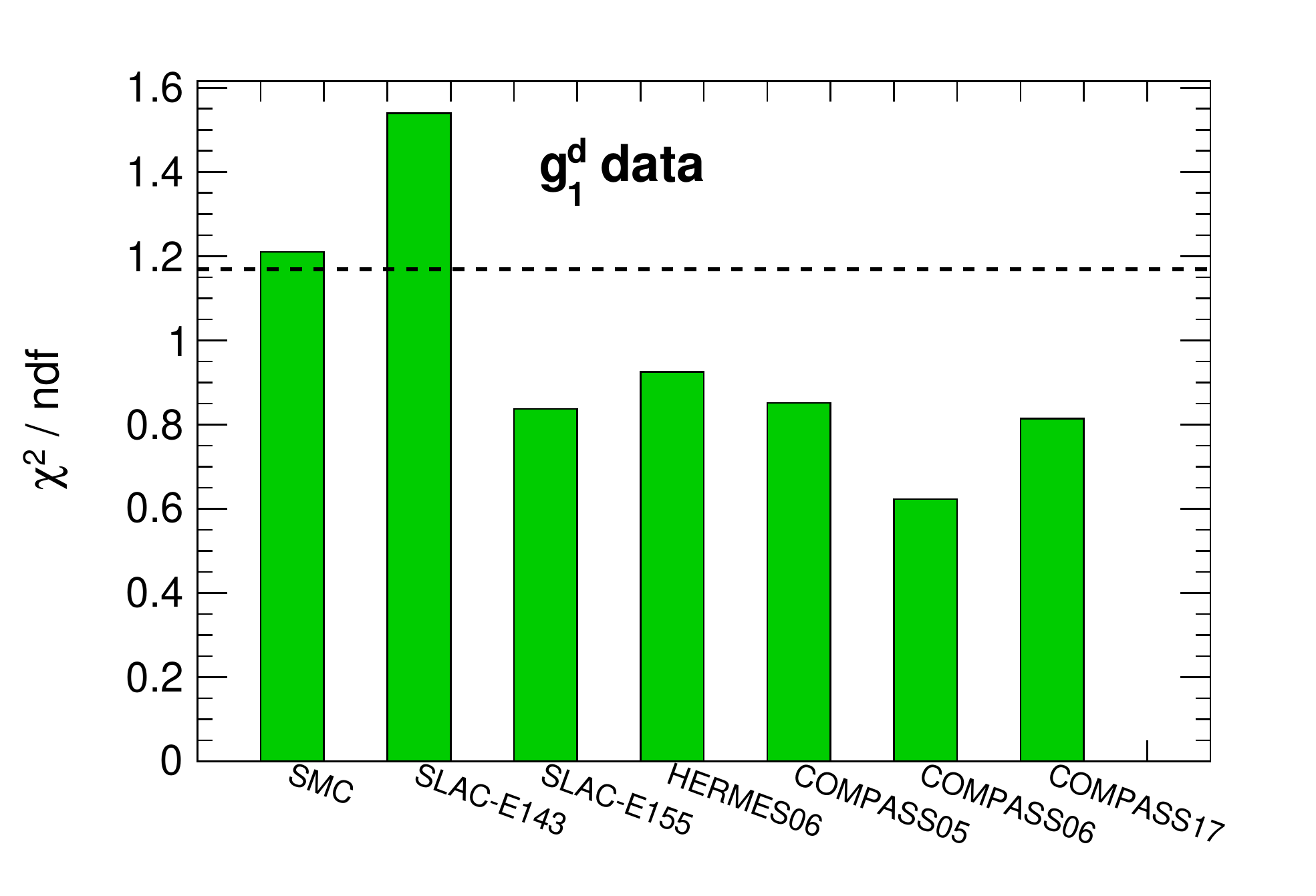}
\caption{
  (color online) Value of the reduced $\chi^{2}$ for the $g_{1}^{d}$ data of each
  experimental group (shown in Fig.\ref{fig:g1d-kinematic-coverage}), included in this global QCD analysis.
  The horizontal black dash line is the weighted average of these reduced $\chi^{2}$ over all the $g_{1}^{d}$ dataset.
}
\label{fig:g1d-chi2}
\end{figure}
\begin{figure}[htbp]
\centering
\includegraphics[scale=0.41]{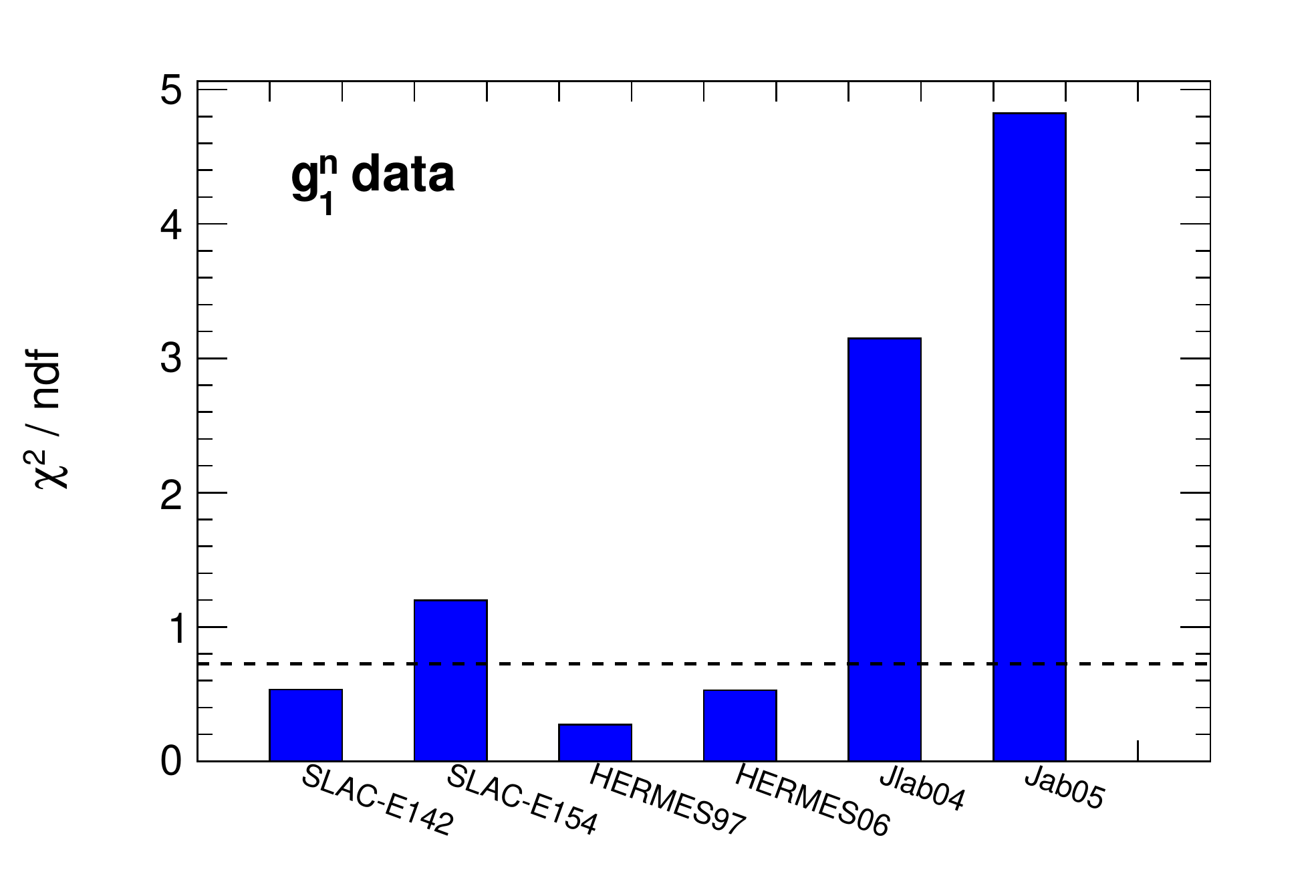}
\caption{
  (color online) Value of the reduced $\chi^{2}$ for the $g_{1}^{n}$ data of each
  experimental group (shown in Fig.\ref{fig:g1n-kinematic-coverage}), included in this global QCD analysis.
  The horizontal black dash line is the weighted average of these reduced $\chi^{2}$ over all the $g_{1}^{n}$ dataset.
}
\label{fig:g1n-chi2}
\end{figure}

Fig.~\ref{fig:g1p-chi2}, Fig.~\ref{fig:g1d-chi2} and Fig.~\ref{fig:g1n-chi2} show the calculated
reduced $\chi^{2}$ values of each experimental group of $g_{1}^{p}$, $g_{1}^{d}$ and $g_{1}^{n}$, respectively.
The horizontal black dashed line represents the weighted average of reduced $\chi^{2}$ of the $g_{1}^{p}$, $g_{1}^{d}$ and $g_{1}^{n}$ datasets,
which are 1.044, 1.169 and 0.726 respectively.
From the distributions of reduced $\chi^{2}$, the weighted average of reduced $\chi^{2}$ of
the experimental data points of $g_{1}^{p}$, $g_{1}^{d}$ and $g_{1}^{n}$ are all less than 1.2,
which shows that the global QCD analysis is fine with the least-square fitting method.
In addition, it can be seen from the figure that whether it is $g_{1}^{p}$ data or $g_{1}^{n}$ data, the value of reduced  $\chi^{2}$ given by the
experimental data of Jlab is larger than that of other experimental cooperation groups.

\begin{table}
  \caption{
    Summary of the obtained values of the reduced $\chi^{2}$ of $g^{p}_{1}$, $g^{d}_{1}$, $g^{n}_{1}$ data sets
    and the number of polarized data points we used for this global QCD analysis.
  }
  \label{tab:data_chi2value_summary}
  \begin{tabular}{cccccccccc}
  \hline\hline
 Observable  & Experiments       &  N points  & $\chi^2/ndf$\\
  \hline
 $g_1^p$ &   EMC + SMC           &    349     & 1.044       \\
         & + SLAC + HERMES       &            &             \\
         & + COMPASS + CLAS      &            &             \\
  \hline
 $g_1^d$ &   COMPASS + HERMES    &    296     & 1.169       \\
         & + SLAC + SMC          &            &             \\
  \hline
 $g_1^n$ &   SLAC + E154         &    118     & 0.726       \\
         & + HERMES + Jlab       &            &             \\
  \hline
  $g_1^p + g_1^d + g_1^n$& Total &    763     & 1.054       \\
  \hline\hline
  \end{tabular}
\end{table}

\begin{table*}
  \caption{
    The obtained input scale $Q_{0}^{2}$ and the parameters of spin-dependent valence quark input (A, B, D, E),
    from the global fit to the worldwide experimental data up to date.
  }
  \label{tab:fit-results}
  \begin{tabular}{cccccccccc}
    \hline\hline
    $Q_0^2$ [GeV$^2$]  &     $A$       &     $B$        &   $D$        & $E$  \\
  \hline
  $0.068\pm0.003$   & $18.85\pm 2.40$ & $3.16\pm 0.17$ & $-6.73\pm 0.57$  & $1.90\pm 0.05$ \\
  \hline\hline
  \end{tabular}
\end{table*}

To study the statistical uncertainties of the polarized PDFs and the related quantities,
the Hessian method \cite{Pumplin:2001ct,Martin:2002aw} is used.
Such statistical uncertainties come from the experimental data of spin structure function
used in this global QCD analysis.
The errors of the free parameters are given with the $\chi^2$-function,
which quantifies the fit between theory and experiment.
Expanding $\chi^2$ around the minimum point $\hat{a}$ and maintaining the leading quadratic term,
we have,
\begin{equation}
\begin{split}
\Delta\chi^2=\chi^2(\hat{a}+\delta a)-\chi^2(\hat{a})=\sum_{i=1}^{n}\sum_{j=1}^{n}H_{ij}\delta a_i \delta a_j,
\end{split}
\label{eq:HessianMatrix}
\end{equation}
where $H_{ij}$ are the elements of Hessian matrix of the parameters,
$a_i$ ($i$ = 1, 2, ..., N) are the model parameters specifying the input polarized PDFs,
and $N$ here is the number of free parameters.

In Hessian matrix method, the uncertainty of any quantity $X$
with respect to the optimized parameters $\hat{a}$ can be calculated using
the standard linear propagation of errors, which is written as,
\begin{equation}
\begin{split}
(\Delta X)^2 = \Delta\chi^2 \sum_{i=1}^{n}\sum_{j=1}^{n} \frac{\partial X}{\partial a_i}C_{ij}(a)\frac{\partial X}{\partial a_j},
\end{split}
\label{eq:ErrorPropagation}
\end{equation}
where $C_{ij}(a)=(H^{-1})_{ij}$ is the second derivative matrix.
The allowed variation $\Delta\chi^2$ of $\chi^2$-function with respect to
the changes of all free parameters depends on the confidence level.
For the parameter space of five dimensions in this analysis,
the allowed $\Delta\chi^2$ is 6.06 at the 70\% confidence level.

\section{Results and discussions}
\label{sec:results}

\begin{figure*}[htbp]
\centering
\includegraphics[scale=0.85]{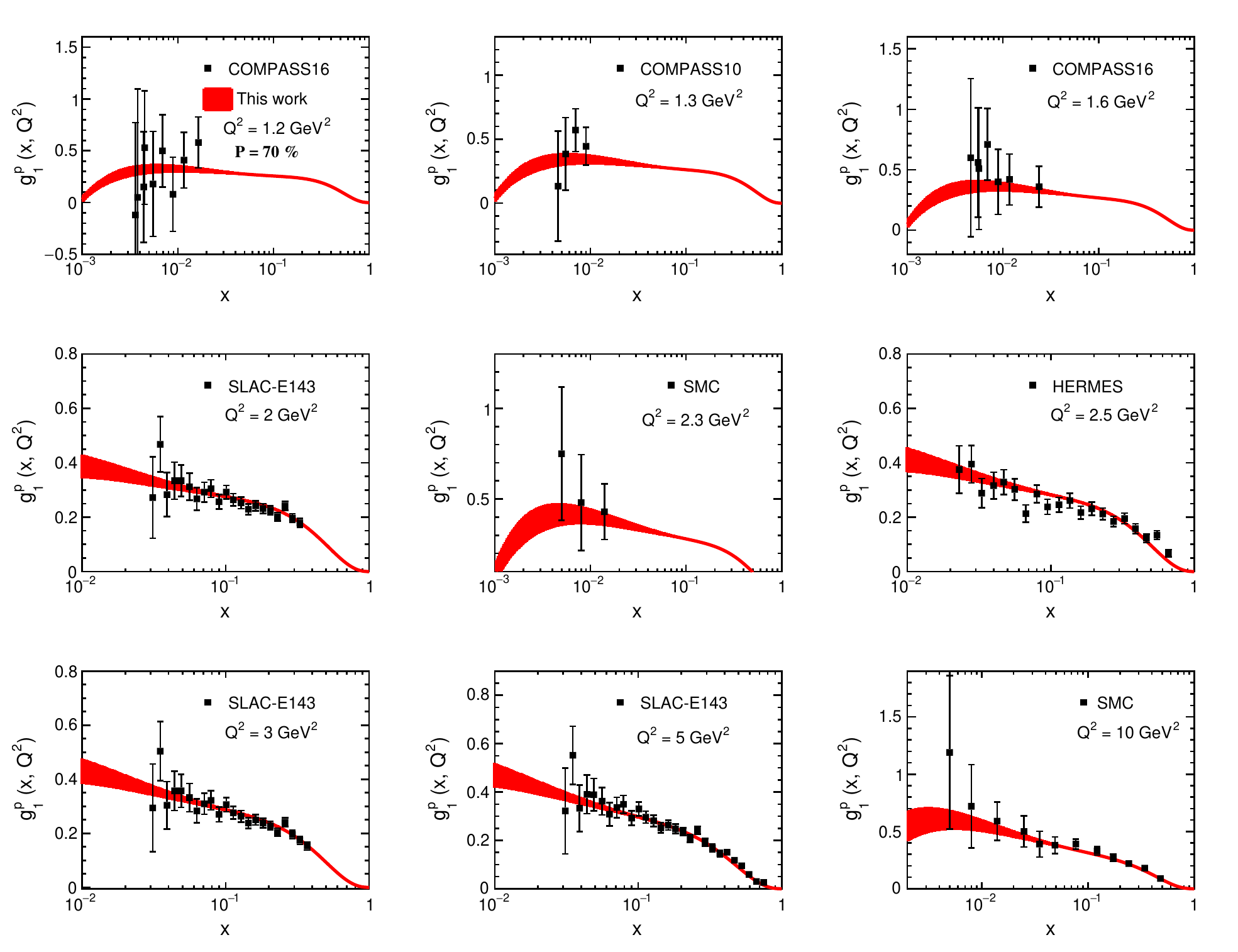}
\caption{(color online) Our predicted polarized structure function $g_{1}^{p}(x, Q^{2})$ as a function of $x$ at different $Q^{2}$
  compared with the polarized DIS data from COMPASS \cite{COMPASS:2010wkz, COMPASS:2015mhb},
  E143 \cite{E143:1998hbs}, SMC \cite{SpinMuon:1998eqa} and HERMES \cite{HERMES:1998cbu, HERMES:2006jyl}.
  The error band of our work shows the 1$\sigma$ statistical uncertainty from Hessian method.  }
\label{fig:g1p-DIS}
\end{figure*}

\begin{figure*}[htbp]
\centering
\includegraphics[scale=0.85]{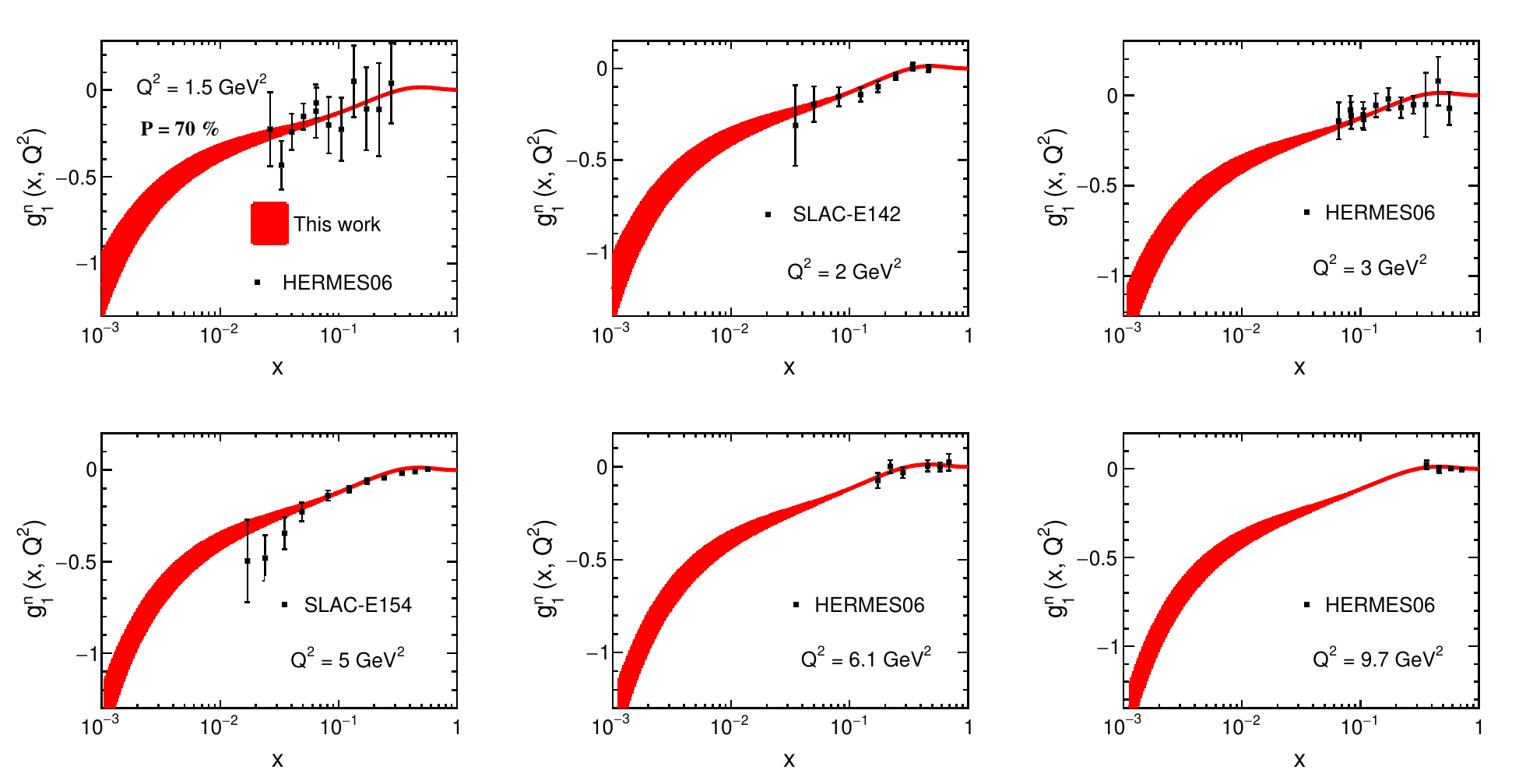}
\caption{(color online) Our predicted polarized structure function $g_{1}^{n}(x, Q^{2})$ as a function of $x$ at different $Q^{2}$
  compared with the polarized DIS data from HERMES \cite{HERMES:2006jyl}, E142 \cite{E142:1996thl} and E154 \cite{E154:1997xfa}.
  The error band of our work shows the 1$\sigma$ statistical uncertainty from Hessian method.}
\label{fig:g1n-DIS}
\end{figure*}

\begin{figure*}[htbp]
\centering
\includegraphics[scale=0.85]{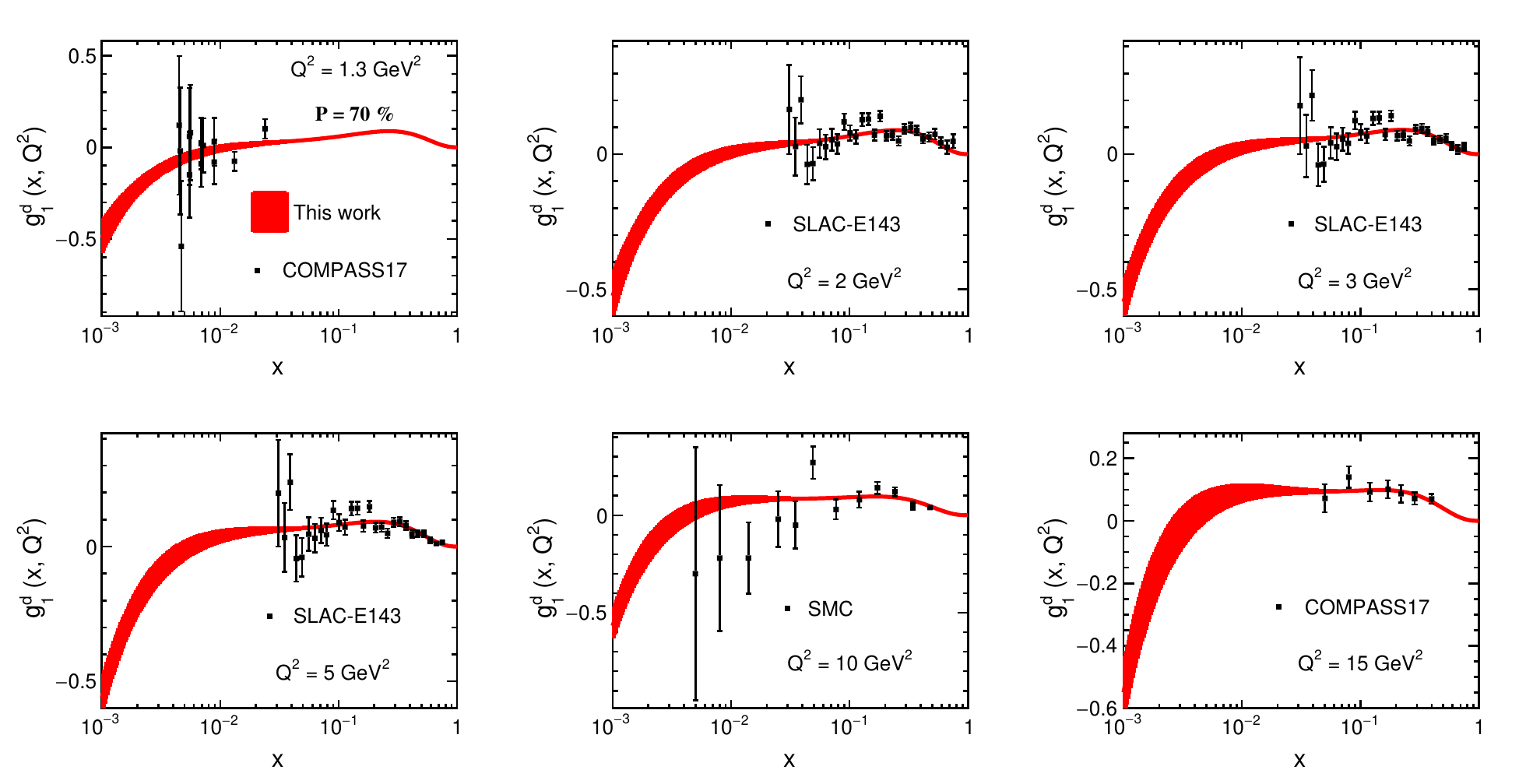}
\caption{(color online) Our predicted polarized structure function $g_{1}^{d}(x, Q^{2})$ as a function of $x$ at different $Q^{2}$
  compared with the polarized DIS data from COMPASS \cite{COMPASS:2016jwv}, E143 \cite{E143:1998hbs} and SMC \cite{SpinMuon:1998eqa}.
  The error band of our work shows the 1$\sigma$ statistical uncertainty from Hessian method.}
\label{fig:g1d-DIS}
\end{figure*}

Fig. \ref{fig:g1p-DIS} shows the comparisons among the obtained polarized structure function $g_{1}^{p} (x, Q^{2})$ of the proton
at various resolution scales $Q^{2} = 1.2, 1.3, 1.6, 2.0, 2.3, 2.5, 3, 5$ and 10 GeV$^{2}$,
with various polarized DIS experiments data from COMPASS \cite{COMPASS:2010wkz, COMPASS:2015mhb},
E143 \cite{E143:1998hbs}, SMC \cite{SpinMuon:1998eqa}, and HERMES \cite{HERMES:1998cbu, HERMES:2006jyl}.
In this figure we find that our predicted polarized structure function $g_{1}^{p} (x, Q^{2})$ are
in agreement with the experimental data over the accessed kinematic region
of $x$ and $Q^2>1$ GeV$^2$.

Fig. \ref{fig:g1n-DIS} shows the comparisons among the obtained polarized structure function $g_{1}^{n} (x, Q^{2})$ of the neutron
at scales $Q^{2} = 2$ and 10 GeV$^{2}$, with polarized DIS experiments data from E142 \cite{E142:1996thl}
and E154 \cite{E154:1997xfa}.
In this figure we find that our predicted polarized structure function $g_{1}^{n} (x, Q^{2})$ are
in agreement with the experimental data over the accessed kinematic region of $x$ and $Q^2>1$ GeV$^2$.

Fig. \ref{fig:g1d-DIS} shows the comparisons among the obtained polarized structure function $g_{1}^{d} (x, Q^{2})$ of the deuteron
at various resolution scales $Q^{2} = 2, 3, 5$ and 10 GeV$^{2}$,
with polarized DIS experiments data from E143 \cite{E143:1998hbs} and SMC \cite{SpinMuon:1998eqa}.
In this figure we find that our predicted polarized structure function $g_{1}^{d} (x, Q^{2})$ are
in agreement with the experimental data over the accessed kinematic region
of $x$ and $Q^2>1$ GeV$^2$.
It is worthy to mention that the current polarized DIS experimental data
in sea-quark region still have big errors and the number of data points is limited.

\begin{figure}[htbp]
\centering
\includegraphics[scale=0.41]{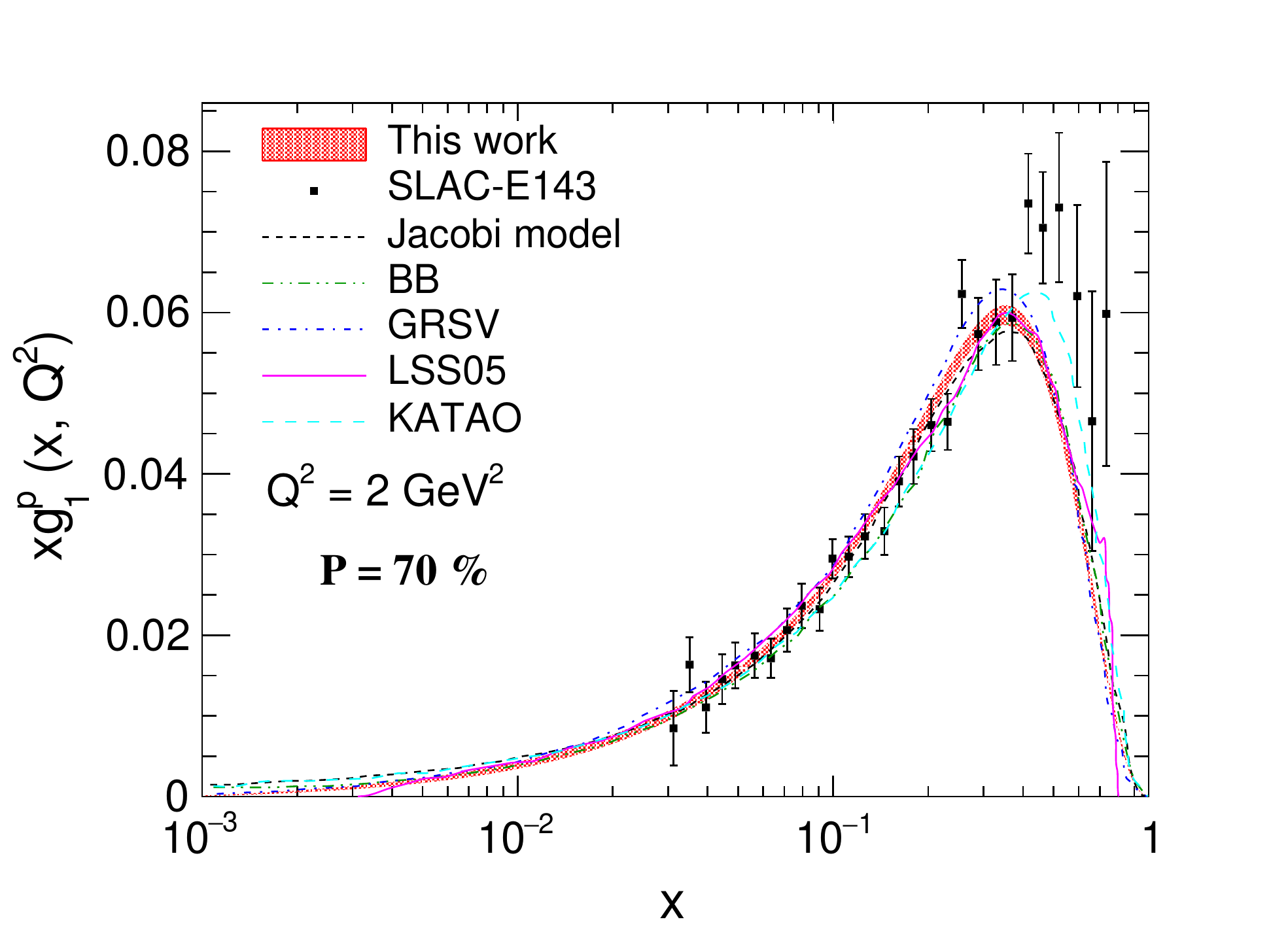}
\caption{
  (color online) Comparisons of our predicted polarized proton structure function $xg_{1}^{p}(x, Q^{2})$ as a function of $x$ at $Q^{2}=2$ GeV$^{2}$
  with the fixed-target E143 experimental data \cite{E143:1998hbs} and some parameterized models, such as Jacobi expansion model
  (short dashed) \cite{Nematollahi:2021ynm}, BB (dashed-dotted-dotted) \cite{Blumlein:2010rn},
  GRSV (dashed-dotted) \cite{Gluck:2000dy}, LSS05 (solid curve) \cite{Leader:2005ci} and KATAO (long dashed) \cite{Khorramian:2010qa}.
  The error band of our work shows the 1$\sigma$ statistical uncertainty from Hessian method.
}
\label{fig:xg1p-E143}
\end{figure}

\begin{figure}[htbp]
\centering
\includegraphics[scale=0.41]{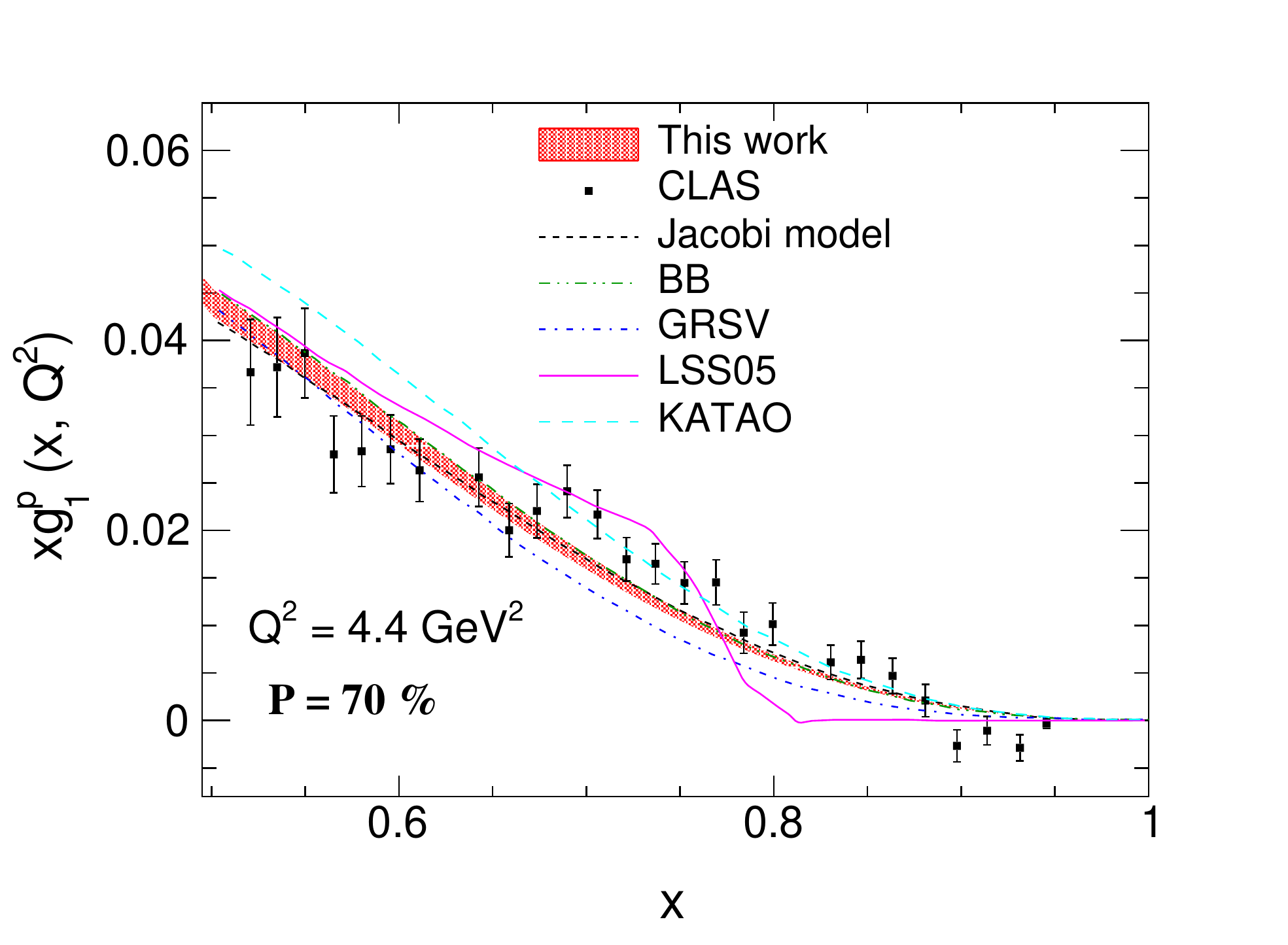}
\caption{
  (color online) Comparisons of our predicted polarized proton structure function $xg_{1}^{p}(x, Q^{2})$ as a function of $x$ at $Q^{2}=4.4$ GeV$^{2}$
  with JLAB17/CLAS experimental data \cite{CLAS:2017qga} and some parameterized models, such as Jacobi expansion model (short dashed) \cite{Nematollahi:2021ynm},
  BB (dashed-dotted-dotted) \cite{Blumlein:2010rn}, GRSV (dashed-dotted) \cite{Gluck:2000dy}, LSS05 (solid curve) \cite{Leader:2005ci} and KATAO (long dashed)
  \cite{Khorramian:2010qa}. The error band of our work shows the 1$\sigma$ statistical uncertainty from Hessian method.
}
\label{fig:xg1p-JLAB17}
\end{figure}

The spin structure function is directly related to the polarized PDFs.
To verify the obtained polarized PDFs and their uncertainties,
we calculate the spin structure function with DGLAP equations with
the nonlinear corrections and the corresponding uncertainties
with the Hessian matrix method.
Fig.~\ref{fig:xg1p-E143} shows our obtained spin structure function $xg_{1}^{p} (x, Q^{2})$
of the proton as a function of $x$ at $Q^{2}=2$ GeV$^{2}$,
which are compared with the E143 experimental data \cite{E143:1998hbs}
and some other phenomenological models,
such as the Jacobi expansion model \cite{Nematollahi:2021ynm},
BB parametrization \cite{Blumlein:2010rn}, GRSV parametrization \cite{Gluck:2000dy},
LSS05 parametrization \cite{Leader:2005ci} and KATAO parametrization \cite{Khorramian:2010qa}.
The error bands in the figure shows the uncertainty at the 70\% of confidence level.
Fig.~\ref{fig:xg1p-JLAB17} shows our obtained spin structure function $xg_{1}^{p} (x, Q^{2})$
of the proton as a function of $x$ at $Q^{2}$ = 4.4 GeV$^{2}$
compared with the recent JLAB17/CLAS experimental data \cite{CLAS:2017qga}
and some other parametrization models.

\begin{figure}[htbp]
\centering
\includegraphics[scale=0.41]{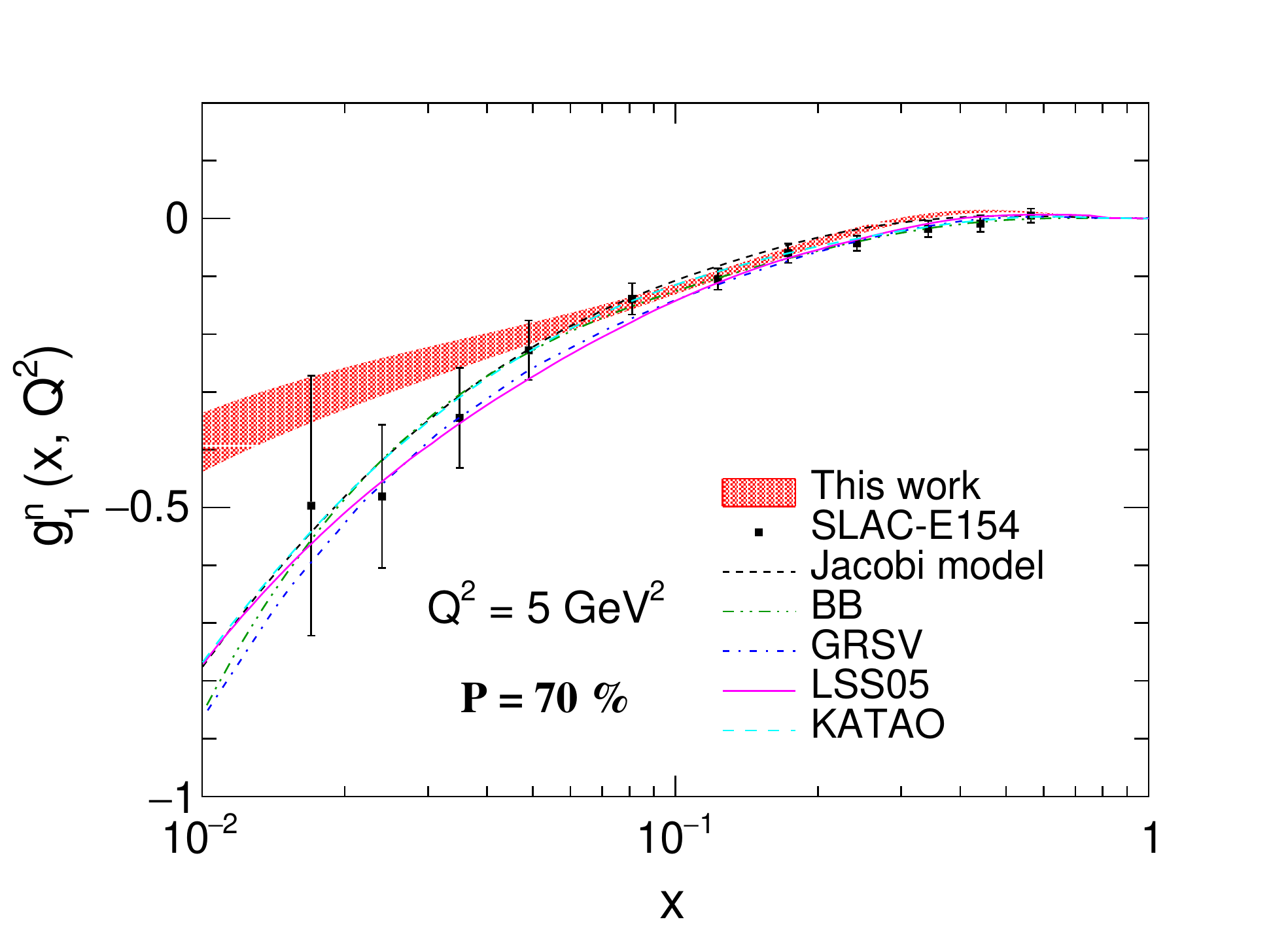}
\caption{
  (color online) Comparisons of our predicted polarized neutron structure function $g_{1}^{n}(x, Q^{2})$ as a function of $x$ at $Q^{2}=5$ GeV$^{2}$
  with the E154 collaboration experimental data \cite{E154:1997xfa} and some parameterized models, such as Jacobi expansion model
  (short dashed) \cite{Nematollahi:2021ynm}, BB (dashed-dotted-dotted) \cite{Blumlein:2010rn},
  GRSV (dashed-dotted) \cite{Gluck:2000dy}, LSS05 (solid curve) \cite{Leader:2005ci} and KATAO (long dashed) \cite{Khorramian:2010qa}.
  The error band of our work shows the 1$\sigma$ statistical uncertainty from Hessian method.
}
\label{fig:g1n-E154}
\end{figure}

Fig.~\ref{fig:g1n-E154} shows our obtained spin structure function $g_{1}^{n} (x, Q^{2})$
of the proton as a function of $x$ at $Q^{2}=2$ GeV$^{2}$,
which are compared with the E154 experimental data \cite{E154:1997xfa}
and some other phenomenological models,
such as the Jacobi expansion model \cite{Nematollahi:2021ynm},
BB parametrization \cite{Blumlein:2010rn}, GRSV parametrization \cite{Gluck:2000dy},
LSS05 parametrization \cite{Leader:2005ci} and KATAO parametrization \cite{Khorramian:2010qa}.
The error bands in the figure shows the uncertainty at the 70\% of confidence level.

\begin{figure}[htbp]
\centering
\includegraphics[scale=0.41]{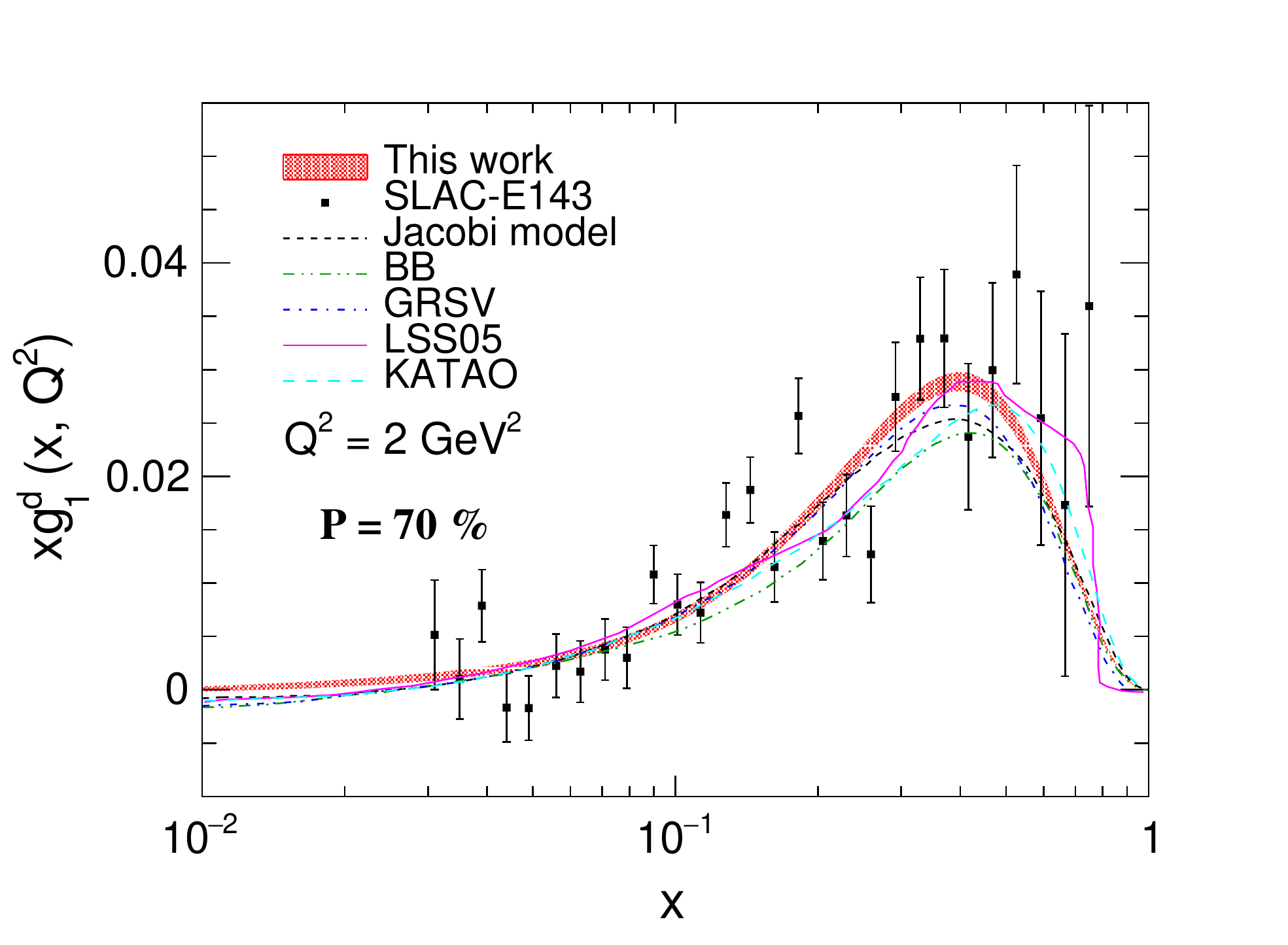}
\caption{
  (color online) Comparisons of our predicted polarized deuteron structure function $xg_{1}^{d}(x, Q^{2})$ as a function of $x$ at $Q^{2}=2$ GeV$^{2}$
  with E143 experimental data \cite{CLAS:2017qga} at SLAC and some parameterized models, such as Jacobi expansion model (short dashed) \cite{Nematollahi:2021ynm},
  BB (dashed-dotted-dotted) \cite{Blumlein:2010rn}, GRSV (dashed-dotted) \cite{Gluck:2000dy}, LSS05 (solid curve) \cite{Leader:2005ci} and KATAO (long dashed)
  \cite{Khorramian:2010qa}. The error band of our work shows the 1$\sigma$ statistical uncertainty from Hessian method.
}
\label{fig:xg1d-E143}
\end{figure}

Fig.~\ref{fig:xg1d-E143} shows our obtained spin structure function $xg_{1}^{d} (x, Q^{2})$
of the deuteron as a function of $x$ at $Q^{2}$ = 2 GeV$^{2}$
compared with the recent E143/SLAC experimental data \cite{CLAS:2017qga}
and some other parametrization models.
We find that the spin structure function of $xg_{1}^{p}$, $xg_{1}^{n}$ and $xg_{1}^{d}$ based on the obtained polarized PDFs
in this work are in agreement with the experimental measurements
in the whole probed region of $x$ up to date.
Our polarized PDFs based on the dynamical parton distribution assumption
are also consistent with the other parameterized models.

\begin{figure*}[htbp]
\centering
\includegraphics[scale=0.85]{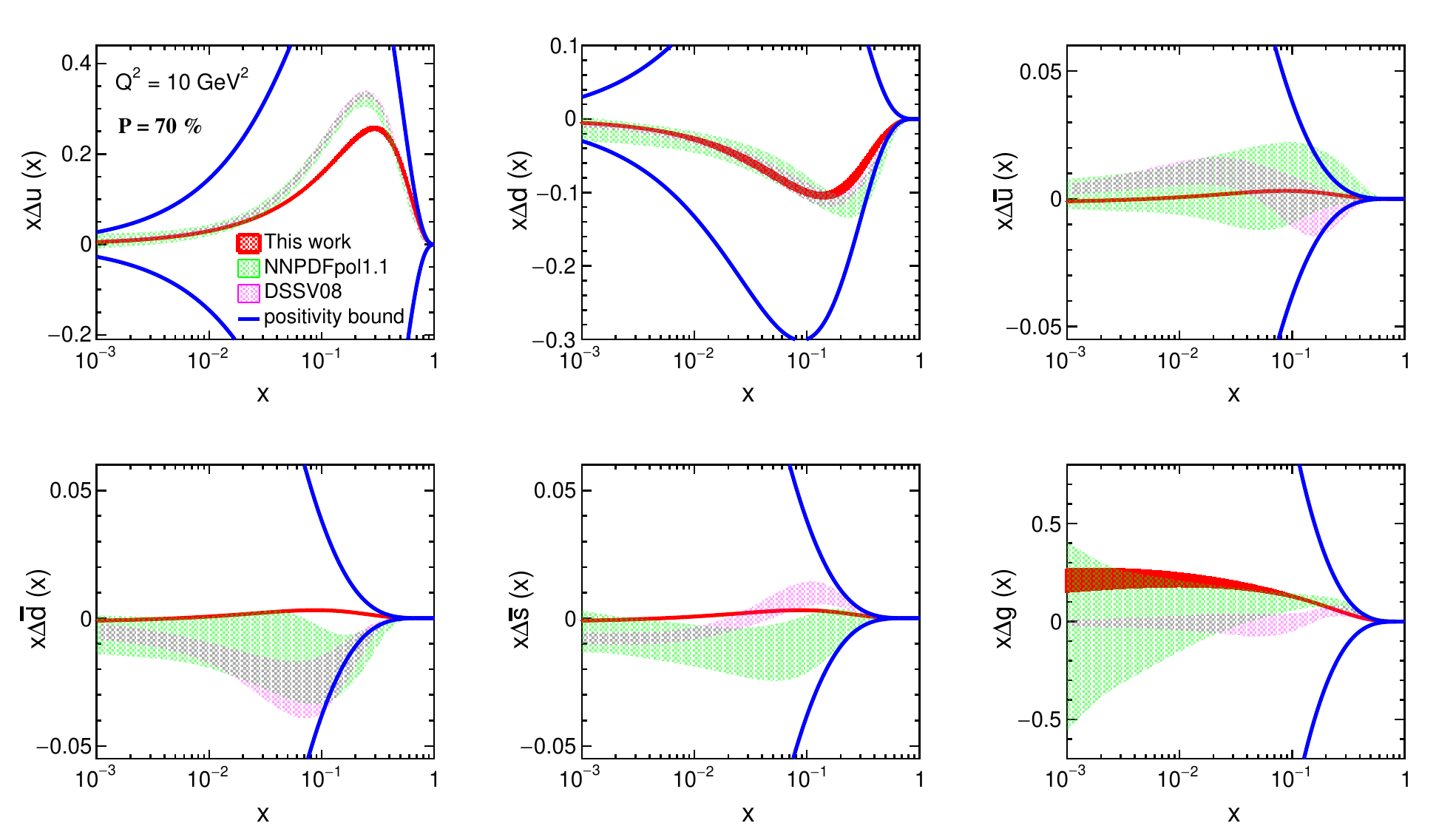}
\caption{(color online) Our predicted polarized PDFs as a function of $x$ at $Q^{2}$ = 10 GeV$^{2}$
  compared with some parameterized models like NNPDF \cite{Nocera:2014gqa} and DSSV08 \cite{deFlorian:2009vb}.
  The error band of our work shows the uncertainty at the 70\% confidence level.   }
\label{fig:pPDFs}
\end{figure*}

Fig.~\ref{fig:pPDFs} shows the obtained polarized valence quark, sea quark and gluon distributions
as a function of $x$ at $Q^2=10$ GeV$^{2}$ from this analysis.
The boundaries of the positive condition are also shown in the figure.
As the cross section is always positive, we have $|\Delta \sigma| < \sigma$
and $-f_i < \Delta f_i < f_i$ at the leading order of the factorization theorem.
We see that for the central value of the obtained polarized PDFs,
the positive condition of the cross section is meet.
Various parametrization models such as NNPDF \cite{Nocera:2014gqa}
and DSSV08 \cite{deFlorian:2009vb} at the $Q^2=10$ GeV$^{2}$
are added for comparisons as well.
The best determined polarized distribution is $x\Delta u_{v}$
from the polarized DIS data on the proton target.
With the addition of the neutron and deuteron DIS data in this analysis,
the uncertainty of the polarized down quark distribution $x\Delta d_{v}$ is limited.
The polarized sea quark distributions are positive and peaked around $x=0.1$,
based on our analysis using the dynamical parton distribution assumptions.
Note that all the sea quark distributions are given by the parton
splitting and recombination processes in the QCD evolution to the high $Q^2$.
Since the flavor-symmetric assumption of sea quarks are used,
the up, down and strange polarized antiquark distributions are identical in this analysis.
The most interesting finding is that the dynamical polarized gluon distribution
is obviously positive, which is consistent with the recent NNPDF analysis
with the jet production data of polarized proton-proton collision.
In general, our obtained polarized PDFs are consistent with the widely used
parameterized PDFs.
With the statistical uncertainties of polarized PDFs estimated,
we see that at small $x$ and the large $x$ ($> 0.8$),
the PDFs are not well constrained by the experimental data.

So far there are still big uncertainties for the polarized gluon and sea quark distributions.
In addition, different global analyses give different predictions on them.
Both NNPDFpol1.1 and our analysis predict that the polarized gluon distribution is positive,
while DSSV08 analysis gives that the polarized gluon distribution is almost zero.
In Fig.~\ref{fig:pPDFs}, we see that on the polarized anti-up quark distribution,
all the parametrization results are more or less consistent with each other.
But on the polarized anti-down and strange quark distributions,
we see the discrepancies.
Our prediction gives the positive $\Delta \bar{d}$,
while NNPDFpol1.1 and DSSV08 give the negative.
Both DSSV08 and our prediction give the positive $\Delta s$,
while NNPDFpol1.1 gives the negative.
These difference on polarized gluon and sea quark distributions
should be solved with the future analysis with much more
experimental data distributed in a much wider kinematic domain.
The polarized electron-ion collider in the future can be employed to this aim.
The inclusive jet and $W^{\pm}$ boson productions from polarized proton collisions
also could be exploited in the analysis to fix the polarized PDFs.

One important part of the spin content of the proton comes from
the spin carried by the individual partons.
The fraction of the proton spin carried by the partons are directly quantified
by the first moments of the polarized PDFs.
In this work, we calculate the truncated first moments of the polarized PDFs
in the region [$x_{\rm min}$, $x_{\rm max}$], which is defined as,
\begin{equation}
\begin{split}
\left<\Delta f(Q^2)\right>^{[x_{\rm min}, ~x_{\rm max}]}
\equiv \int_{x_{\rm min}}^{x_{\rm max}} dx \Delta f(x, Q^2).
\end{split}
\label{eq:1stMomentDef}
\end{equation}
We also will provide the full moments, $\left<\Delta f(Q^2)\right>^{[0,1]}$.
The first moments of different types of quarks and the sum are listed in Table \ref{tab:QuarkFirstMoment}.
Here the sum is defined as $\Delta \Sigma = \Delta u + \Delta \bar{u} + \Delta d + \Delta \bar{d} + \Delta s + \Delta \bar{s}$,
ignoring the contributions from heavy quarks which are tiny
compared to that from light quarks.
We also define the $C$-even combinations as,
$\Delta u^{+} = \Delta u + \Delta \bar{u}$ and $\Delta d^{+} = \Delta d + \Delta \bar{d}$.
The truncated and full moments of the gluon polarized distribution is listed in Table \ref{tab:GluonFirstMoment}.

For the truncated first moment of all quarks $\left<\Delta \Sigma \right>^{[10^{-3},1]}$, our prediction is close to
the value DSSV08 ($0.366_{-0.062}^{+0.042}$) while larger than the value of NNPDFpol1.1 ($0.25\pm 0.10$).
For the full moment $\left<\Delta \Sigma \right>^{[0,1]}$,
our prediction is obviously larger than the NNPDFpol1.1's prediction of $0.18\pm 0.21$.
This is due to the different extrapolations to the small $x$.
According to our calculation, the spin of all the quarks contribute
about 30$\%$ of the proton spin at $Q^2=10$ GeV$^2$.
In Table \ref{tab:GluonFirstMoment}, we find that for the truncated gluon moment $\left<\Delta g \right>^{[0.05,0.2]}$,
our calculation results are approximately consistent with NNPDF and DSSV++
However for the full gluon moment $\left<\Delta g \right>^{[0,1]}$,
our prediction of gluon spin is very big. This implies that there is a significant negative orbital angular
momentum of the gluons, which can be studied with the gluon generalized parton
distribution in hard exclusive electroproduction channel of J/$\psi$ vector meson.

\begin{table}[h]
	\caption{\label{tab:QuarkFirstMoment}
             The truncated and full moments of the quarks of different flavors
             and the sum of all of them, from this work.
             The $Q^2$ is at 10 GeV$^2$.
             The uncertainties are 1$\sigma$ statistical errors derived from experimental data.    }
	\begin{tabular}{ccc}
            \hline\hline
            $\delta f$        & $\left<\delta f(Q^2)\right>^{[0,~1]}$ & $\left<\delta f(Q^2)\right>^{[10^{-3},~1]}$ \\
            \hline
            $\delta u^{+}$    & $0.622\pm 0.012$      & $0.615\pm 0.014$       \\
            $\delta d^{+}$    & $-0.312\pm 0.017$     & $-0.287\pm 0.014$      \\
            $\delta \bar{u}$  & $-0.0022\pm 0.0005$   & $0.0070\pm 0.0024$   \\
            $\delta \bar{d}$  & $-0.0022\pm 0.0005$   & $0.0070\pm 0.0024$   \\
            $\delta s$        & $-0.0022\pm 0.0005$   & $0.0070\pm 0.0024$   \\
            $\delta \Sigma$   & $0.301\pm 0.014$      & $0.356\pm 0.015$      \\
            \hline\hline
	\end{tabular}
\end{table}

\begin{table}[h]
	\caption{\label{tab:GluonFirstMoment}
             The truncated and full moments of the gluon polarized distributions from
             various global analyses.
             The $Q^2$ is at 10 GeV$^2$.
             The uncertainties are 1$\sigma$ statistical errors derived from experimental data.   }
	\begin{tabular}{cccc}
            \hline\hline
            PDF set      & $\left<\delta f(Q^2)\right>^{[0,1]}$ & $\left<\delta f(Q^2)\right>^{[10^{-3},1]}$ & $\left<\delta f(Q^2)\right>^{[0.05,0.2]}$ \\
            \hline
            This work    &  $2.09\pm 0.64$   &    $1.02\pm 0.18$     &   $0.17\pm0.02$    \\
            NNPDFpol1.0  &  $-0.95\pm 3.87$  &    $-0.06\pm 1.12$   &    $0.05\pm0.15$    \\
            NNPDFpol1.1  &  $0.03\pm 3.24$   &    $0.49\pm 0.75$    &    $0.17\pm0.06$    \\
            DSSV08       &          -        &    $0.13_{-0.314}^{+0.702}$    &    $0.005_{-0.164}^{+0.129}$    \\
            DSSV++       &          -        &           -          &    $0.10_{-0.07}^{+0.06}$    \\
            \hline\hline
	\end{tabular}
\end{table}

\section{Summary}
\label{sec:summary}

In this work, we studied a simple parametrization for the initial polarized parton distributions
and have determined the polarized PDFs from the QCD analysis of DIS data.
The corresponding statistical uncertainties of the polarized PDFs
are provided with the Hessian method.
It is interesting to find that the polarized gluon distribution is quite positive
in the whole $x$ region.
The other interesting finding is that the polarized sea quark distribution is small
however not zero.
All sea quarks and gluons are dynamically generated in the QCD evolution process.
These dynamically produced parton distributions give us some important knowledge
on the polarized parton distributions,
and they help us better understand the proton spin decomposition.
The obtained polarized PDFs are compared with the analyses from other groups,
and the consistences are found within the statistical errors.

We have obtained the corresponding hadronic scale $Q_0^2$
for the naive nonperturbative input of only three valence quarks.
The obtained input scale is around 0.68 GeV$^2$,
which is very close to the previous value determined from the unpolarized DIS data \cite{Wang:2016sfq}.
The parametrization of the input valence quark distributions
are also determined.
The polarized PDFs from this nonperturbative input reproduce well
the spin structure function measured at high $Q^2$,
which implies that the quark model input provides a roughly
good origin of the polarized PDFs.
The previous parton correlation length $R$ and the saturated strong running
coupling work well for the polarized DGLAP equations with parton-parton recombination corrections.
The QCD parameters and the parton-parton recombination corrections
QCD evolution equations are a simple and valid bridge
to connect the pure valence nonperturbative input with the experimental measurements at the hard scales.
The global QCD analysis based on the pure dynamical parton model clearly shows
how the parton distributions are generated.

The obtained polarized PDFs of the proton from this analysis
are suggested to be used in a lot of phenomenological applications.
Nevertheless we also see that there are some rooms left to be improved in the future,
such as using more accurate proton, neutron and deuteron experimental data for the global analysis,
taking into account the next-to-leading order corrections to the DGLAP equations with the parton-parton recombination effect,
including the experimental data of inclusive jet or identified hadron or W boson production of polarized proton collisions.
Most importantly, much more polarized DIS experimental data of the proton and the deuteron
should be acquired to fix the polarized sea quark and gluon distributions precisely.
In the future, the polarized DIS data in a much wide kinematical region will be obtained from facilities
such as electron-ion collider in US \cite{Accardi:2012qut,AbdulKhalek:2021gbh} 
and China \cite{Chen:2020ijn,Chen:2018wyz,Anderle:2021wcy}
with the polarized beams for $e-p$ collisions.

\begin{acknowledgments}
We thank professors Wei Zhu, Fan Wang and Jianhong Ruan for the discussions and suggestions.
This work is supported by the Strategic Priority Research Program of Chinese Academy of Sciences under the Grant NO. XDB34030301,
the National Natural Science Foundation of China under the Grant NO. 12005266
and the Guangdong Major Project of Basic and Applied Basic Research under the Grant No. 2020B0301030008.
\end{acknowledgments}

\bibliographystyle{apsrev4-1}
\bibliography{refs}

\end{document}